\PassOptionsToPackage{table}{xcolor}
\documentclass[10pt,sigconf,letterpaper,nonacm]{acmart}

\usepackage{array}
\usepackage{booktabs}
\usepackage{graphicx}
\usepackage{pifont}
\usepackage[caption=false,font=footnotesize]{subfig}
\usepackage{url}
\usepackage{enumitem}

\setlist[itemize,enumerate]{wide=4pt}

\newcommand{\yesmark}{\ding{51}}
\newcommand{\nomark}{\ding{55}}

\title{Beyond the Next Hop: Measuring Data-Plane Forwarding for RPKI-Invalid Prefixes in the Wild}

\newcommand{\tsinghua}{Tsinghua University}
\newcommand{\pekinguniversity}{Peking University}
\newcommand{\zhongguancun}{Zhongguancun Laboratory}

\makeatletter
\newcommand{\paperbyline}{}
\newcommand{\paperauthor}[4][]{%
  \author{#2}%
  \ifstrempty{#1}{\def\paperauthormark{}}{%
    \authornote{#1}\def\paperauthormark{*}}%
  \affiliation{\institution{#4}\city{Beijing}\country{China}}%
  \ifdefempty{\paperbyline}{}{\gappto\paperbyline{, }}%
  \eappto\paperbyline{\noexpand\mbox{#2\noexpand\textsuperscript{#3\paperauthormark}}}%
}
\newcommand{\compactpaperauthors}{%
  \gdef\@currentauthors{}%
  \global\setbox\mktitle@bx=\vbox{%
    \unvbox\mktitle@bx\par\medskip
    {\centering\@authorfont\paperbyline\par}%
    {\centering\@affiliationfont
      \textsuperscript{1}\tsinghua\qquad
      \textsuperscript{2}\pekinguniversity\qquad
      \textsuperscript{3}\zhongguancun\par}%
    \bigskip
  }%
}
\if@ACM@nonacm
  \if@ACM@anonymous\else
    \let\@mkauthors@iii\compactpaperauthors
  \fi
\fi
\makeatother

\paperauthor{Renrui Tian}{1}{\tsinghua}
\paperauthor[Yahui Li is the corresponding author of this paper.]{Yahui Li}{1}{\tsinghua}
\paperauthor{Xingang Shi}{1}{\tsinghua}
\paperauthor{Xiao Liu}{2}{\pekinguniversity}
\paperauthor{Ying Tian}{3}{\zhongguancun}
\paperauthor{Yanbo Yin}{1}{\tsinghua}
\paperauthor{Xinzhe Liu}{1}{\tsinghua}
\paperauthor{Yuhang Li}{1}{\tsinghua}
\paperauthor{Jiarui Li}{1}{\tsinghua}
\paperauthor{Yujie Huang}{1}{\tsinghua}
\paperauthor{Han Zhang}{1}{\tsinghua}
\paperauthor{Zhiliang Wang}{1}{\tsinghua}
\paperauthor{Xia Yin}{1}{\tsinghua}

\begin{document}

\setlength{\emergencystretch}{2em}

\begin{abstract}
An operator can inspect local routing and forwarding tables to identify the next hop for a
destination, but cannot tell toward which origin traffic is forwarded beyond that hop. We study
this question for IPv6 prefixes with RPKI-invalid announcements in the wild. We develop a
context-aware path association method that derives discriminative features near candidate origins
from collector-visible BGP paths and uses them to infer forwarding directions from incomplete
traceroutes. Across 82 measurement rounds, we collected 169.5 million traceroutes from 1,811 ASes
toward 2,764 target prefixes.

Our measurements show that traffic is less often forwarded toward invalid origins when non-invalid
alternatives are available, with fewer than 10\% of traceroutes showing forwarding exclusively toward invalid
origins, compared with about a quarter without alternatives. Notably, when alternatives are absent,
we find that
prefixes with identifiable operational explanations retain substantially broader reachability
than unexplained prefixes, suggesting that connectivity considerations may drive continued
invalid-route propagation. Our case analyses further reveal how interconnection, filtering, and
export policies shape the propagation and use of invalid routes, including collateral damage under
partial ROV deployment. By looking beyond the locally selected next hop, our work bridges the gap
between local route selection and actual downstream forwarding, uncovering operational blind spots
that local routing data alone cannot reveal.
\end{abstract}

\maketitle

\section{Introduction}

The Internet consists of autonomous systems (ASes) that use the Border Gateway Protocol (BGP) to
exchange routing information for IP prefixes.  Because BGP does not verify whether an AS is
authorized to originate a prefix, interdomain routing remains vulnerable to accidental
misconfigurations and malicious prefix hijacks
~\cite{lad2007resiliency,sermpezis2018survey,cho2019classification,ripe2008youtube,secureworks2014bitcoin,amazonsoc2018route53,kentik2023bgpincidents}.
The Resource Public Key Infrastructure (RPKI)~\cite{rfc6480}
provides cryptographically verifiable Route Origin Authorizations (ROAs)~\cite{rfc9582}, which bind prefixes and
permitted prefix lengths to authorized origin ASes.  Networks can
use these authorizations to perform Route Origin Validation (ROV)~\cite{rfc6811,rfc8481} and assign a validation state to
each BGP announcement.  Whether and how an AS incorporates that state into its
routing decisions, however, is determined by local policy: a network may reject an invalid
announcement, reduce its preference, or continue to accept it.  Despite partial visibility from
operator disclosures and public routing data, local routing policies and route selection remain
difficult to observe systematically at Internet scale.

There have been two main lines of measurement efforts to understand how networks handle
RPKI-invalid announcements.  Controlled experiments manipulate the validation state of
announcements for researcher-operated prefixes and compare the resulting control-plane and
data-plane observations to isolate routing changes attributable to ROV filtering
~\cite{reuter2018rigorous,rodday2021revisiting,hlavacek2023routeservers,li2025improv,li2026rscope}.
In-the-wild measurements instead investigate naturally occurring invalid announcements.  Some
infer ROV filtering or protection from active reachability and path
measurements~\cite{chen2022rovmi,li2023rovista}, while other work examines a broader set of invalid
announcements and characterizes their associated reachability loss and path
divergence~\cite{li2026demystifying}.

While previous studies have substantially advanced our understanding of ROV in practice, a
practical concern for network operators remains underexplored: \emph{when my AS sends traffic to a
prefix with an RPKI-invalid announcement, toward which origin is it forwarded after the next hop?}
Based solely on local routing and forwarding information, an operator cannot determine the routing
decisions made by other on-path networks.

Measuring this forwarding direction poses two challenges. First, data-plane measurements
face a scarcity of measurable targets. As ROV deployment expands, invalid routes propagate
to fewer networks, leaving fewer vantage points with paths via these
routes~\cite{li2026rscope}. Even when a prefix remains reachable, ICMP filtering can suppress
probe responses~\cite{rfc4890}. IPv6 further complicates active probing: its vast, sparsely
populated address space hampers responsive target discovery~\cite{gasser2018clusters,beholder2018ipv6},
temporary addressing makes discovered targets short-lived~\cite{rfc8981}, and ICMPv6 rate limiting
further suppresses replies~\cite{rfc4443,beholder2018ipv6}.

Second, connectivity tests alone cannot resolve origin ambiguity. A response can be attributed
to a specific origin only when all routes covering the target address originate from that AS.
When different ASes originate the same prefix or overlapping prefixes covering the target
address, a response confirms reachability but cannot identify the origin toward which traffic
was forwarded.

In this paper, we present a large-scale measurement study of data-plane forwarding beyond the
next hop for prefixes with naturally occurring RPKI-invalid IPv6 announcements. We construct
collector-visible routing contexts from contemporaneous public BGP and RPKI data and collect
distributed IPv6 traceroutes through open 6in4 tunnel endpoints. Our context-aware path
association method uses origin-proximal AS nodes, links, and ordered AS pairs to infer
forwarding directions from incomplete observed paths.

Our measurement campaign collected 169.5 million traceroutes toward 2,764 target prefixes from
1,811 ASes across 82 rounds. We use 2,077,525 traceroutes from a complete measurement round for
cross-sectional analysis and all 82 rounds for longitudinal analysis. Our contributions and main
findings are:
\begin{itemize}[topsep=8pt]
  \item \emph{Context-Aware Path Association.} We develop a context-aware path association method
        that uses origin-proximal features to infer forwarding directions from incomplete
        traceroutes. AS links and ordered AS pairs distinguish candidate paths that share
        individual ASes. Our evaluation demonstrates accurate forwarding inference and recovery
        under response loss, as well as resilience to limited control-plane visibility and temporal offsets.
  \item \emph{Forwarding Analysis.} We characterize forwarding directions, prevalence, and
        persistence across routing contexts. When non-invalid alternatives coexist,
        88.6--97.2\% of traceroutes with a unique forwarding direction are forwarded toward
        alternative origins. Although invalid-origin forwarding spans most measured prefixes and nearly
        all measured ASes, it involves fewer ASes per prefix when alternatives exist.
        Longitudinal measurements further reveal that forwarding observations generally remain
        stable under unchanged routing contexts and that invalid-origin forwarding is often
        observed on multiple measurement days for the same AS--prefix pair.
  \item \emph{Operational Insights.} We find that prefixes with operational explanations
        retain substantially broader reachability than unexplained prefixes when no alternative
        is available, suggesting that connectivity considerations may drive continued
        invalid-route propagation. Our analyses reveal how interconnection topology, route
        filtering, and export policies shape the propagation and use of invalid routes. We also
        uncover a case of collateral damage under partial ROV deployment.
\end{itemize}

We will make our measurement and analysis datasets publicly available to support reproducibility
and further research.

\section{Background and Related Work}
\label{sec:background}

\subsection{BGP and RPKI}

BGP enables ASes to exchange reachability information and establish interdomain routes.  A BGP
announcement associates an IP prefix with an AS path, whose final AS identifies the route origin.
When an AS learns multiple routes to the same prefix, it applies its local routing policy to select
a best route.  However, BGP itself does not verify whether the origin AS is authorized to announce
the prefix~\cite{rfc4271}.

RPKI complements BGP by providing cryptographically verifiable authorizations for route origins.
A ROA~\cite{rfc9582} is a signed RPKI object that authorizes an AS to originate one or more IP prefixes.  The
authorization is issued under an RPKI certificate covering the relevant address resources.  For
each prefix, the ROA may include an optional \emph{maxLength} that specifies the longest prefix
length the AS is authorized to announce.  The ROA and the supporting objects required to validate
it are published at an RPKI publication point~\cite{rfc6480}.

To perform ROV, network operators use Relying Party (RP) software to retrieve RPKI objects from
publication points and cryptographically validate them.  From the validated ROAs, the RP derives a
set of Validated ROA Payloads (VRPs), each specifying an authorized prefix, an origin AS, and a
maximum prefix length.  The RP provides these VRPs to routers through the RPKI-to-Router (RTR)
protocol~\cite{rfc8210}.

For each BGP route, a router compares the route prefix and origin AS against its local set of VRPs.
It first determines whether the route prefix is \emph{covered} by any VRP---that is, whether the
route prefix is identical to or more specific than the VRP prefix.  A covering VRP \emph{matches}
the route if it also authorizes the route's origin AS and permits its prefix length.  Based on
these comparisons, the router assigns one of three validation
states~\cite{rfc6811,rfc8481}:

\begin{itemize}
  \item \emph{Valid.}
        At least one VRP matches the route.

  \item \emph{Invalid.}
        At least one VRP covers the route prefix, but none matches the route.

  \item \emph{Unknown (not-found).}
        No VRP covers the route prefix.
\end{itemize}

Operators can incorporate validation states into routing decisions by rejecting invalid
routes~\cite{rfc6811,rfc8481}.  However, recent studies show that many invalid routes result from
ROA misconfigurations or legitimate operational practices rather than prefix
hijacks~\cite{chen2025rightship,li2026demystifying}.

In this paper, we use \emph{non-invalid} to refer collectively to valid and unknown states.

\begin{table}[ht]
  \centering
  \caption{Route categories in the four routing contexts.}
  \label{tab:contexts}
  \footnotesize
  \begin{tabular}{@{}lccc@{}}
    \toprule
      & \multicolumn{2}{c}{\textbf{Exact-prefix}}
      & \textbf{Less-specific} \\
    \cmidrule(lr){2-3}\cmidrule(l){4-4}
    \textbf{Routing context}
      & \textbf{Invalid}
      & \textbf{Valid}
      & \textbf{Non-invalid} \\
    \midrule
    None          & \yesmark & \nomark  & \nomark  \\
    Exact-prefix  & \yesmark & \yesmark & \nomark  \\
    Less-specific & \yesmark & \nomark  & \yesmark \\
    Both          & \yesmark & \yesmark & \yesmark \\
    \bottomrule
  \end{tabular}
\end{table}

\subsection{Routing Contexts}

An RPKI-invalid announcement may coexist with non-invalid same-prefix or less-specific routes.  If
a network filters the invalid route, the prefix may become unreachable when no alternative is
available, or remain reachable through an alternative.  If the invalid route is retained, it may
still be selected.  We therefore interpret invalid announcements together with their coexisting
routes rather than in isolation.

We use \emph{routing context} to denote the combination of route categories that coexist in the
control plane for a prefix.  Table~\ref{tab:contexts} summarizes the four contexts considered in
this work.

\begin{enumerate}
  \item \emph{No alternative.}
        Only invalid routes to the prefix are present, with no non-invalid alternative routes.

  \item \emph{Exact-prefix alternative.}
        The prefix is observed as a multiple-origin AS (MOAS) prefix in which invalid and valid
        routes coexist.  A network that learns both valid and invalid routes may select either
        according to its local routing policy.

  \item \emph{Less-specific alternative.}
        Invalid routes to the prefix coexist with one or more non-invalid routes to less-specific
        prefixes.  If an invalid route is installed, longest-prefix matching selects it for
        forwarding; otherwise, traffic may be forwarded via a less-specific alternative.

  \item \emph{Both alternatives.}
        Both exact-prefix and less-specific alternatives coexist with the invalid routes.  A
        network may select a valid or invalid exact-prefix route according to its local routing
        policy; if no exact-prefix route is installed, traffic may instead be forwarded via a
        less-specific alternative.
\end{enumerate}

\subsection{Related Work}

\paragraph{Controlled experiments}
Controlled experiments manipulate the RPKI validation state of announcements for
researcher-controlled prefixes, allowing observed differences in control-plane propagation and
data-plane behavior to be attributed more directly to ROV filtering
~\cite{reuter2018rigorous,rodday2021revisiting,hlavacek2023routeservers,li2025improv,li2026rscope,
hlavacek2018methodologies,cloudflare2020safe,apnic2021rov}. Control-plane studies typically use
public BGP collectors to examine route propagation
~\cite{reuter2018rigorous,hlavacek2018methodologies}, while data-plane measurements analyze
forwarding paths through RIPE Atlas traceroutes
~\cite{rodday2021revisiting,hlavacek2018methodologies}, test reachability through TCP connections
~\cite{hlavacek2018methodologies}, or assess user-visible protection through browser-based HTTP
requests~\cite{cloudflare2020safe,apnic2021rov}. Beyond these efforts, Hlavacek et
al.~\cite{hlavacek2023routeservers} introduce an AS classification scheme that incorporates path
divergence to refine ROV deployment inference; Li et al.~\cite{li2025improv} coordinate exact and
less-specific announcements to investigate collateral effects on reachability; and Li et
al.~\cite{li2026rscope} selectively serve different ROAs to distinguish local filtering from
upstream protection and identify dependencies on RPKI validators. By design, however, these experiments are not intended to characterize invalid announcements in the wild.

\paragraph{In-the-wild measurements}
Another thread of work concerns naturally occurring RPKI-invalid announcements
~\cite{reuter2018rigorous,chen2022rovmi,li2023rovista,li2026demystifying,tian2025eris,
gilad2016deployment,testart2020filter,chen2025rightship}. Control-plane studies
analyze in-the-wild invalid routes using public BGP data
~\cite{reuter2018rigorous,gilad2016deployment,testart2020filter}, yet limited visibility into
routing decisions can lead to misclassification~\cite{reuter2018rigorous}. ROV-MI collects
traceroute paths toward invalid prefixes and applies Bayesian inference to estimate AS-level ROV
deployment~\cite{chen2022rovmi}, whereas RoVista uses an IP-ID side channel to infer remote
reachability~\cite{li2023rovista}. Both methods restrict their measurements to prefixes
without coexisting valid same-prefix or non-invalid less-specific alternatives and
require suitable measurement targets within those prefixes. Consequently, they do not characterize
forwarding behavior in routing contexts where invalid announcements coexist with such alternatives.
Complementary work assesses the legitimacy of in-the-wild invalid
announcements~\cite{chen2025rightship} and investigates their operational causes and data-plane
impact by measuring reachability loss and path divergence~\cite{li2026demystifying}.

\section{Measurement Methodology}
\label{sec:method}

We conduct a large-scale Internet measurement campaign to characterize data-plane forwarding
directions for prefixes with RPKI-invalid announcements.  We first identify target prefixes and
construct their collector-visible routing contexts using public BGP and RPKI data.  We then use
open 6in4 tunnel endpoints as distributed vantage points to collect IPv6 traceroutes toward the
selected prefixes.  Finally, we relate the observed paths to the candidate origins represented in
their routing contexts.

\subsection{Target Prefix Selection and Routing Contexts}
\label{sec:target-selection}

We use public BGP data to select target prefixes and construct their collector-visible routing
contexts.  Every 2 hours, we obtain IPv6 RIB dumps from RouteViews~\cite{routeviews} and
validated ROA payloads (VRPs) from Routinator~\cite{routinator}.  We remove routes associated
with reserved prefixes or private-use ASNs and validate the remaining routes against these VRPs.
To mitigate biases from local policies and transient routing conditions, we retain only
prefixes whose invalid routes are observed by at least three collectors.  For each retained
prefix, we identify coexisting valid exact-prefix routes and non-invalid less-specific
alternatives.  We exclude less-specific routes shorter than /13 because their broad address
coverage and limited visibility could contaminate the resulting routing contexts.

In examining RPKI-invalid prefixes observed on June 30, 2026, we find that for 2,562 of the
3,838 prefixes (66.8\%), invalid exact-prefix routes and non-invalid less-specific alternatives
are originated by the same AS.  For these prefixes, when invalid routes are filtered,
less-specific alternatives can preserve reachability to the same origin AS, although the resulting
forwarding paths may diverge.  Prior work shows that such path divergence can result in suboptimal
routing, including increased hop count and latency or the bypass of upstream ASes that provide
transit or security services~\cite{li2026demystifying}.  Since our study focuses on forwarding
directions toward different origin ASes, we exclude these prefixes from subsequent traceroute
measurements.

Using public BGP data, we construct the collector-visible routing context of each remaining prefix
from its invalid exact-prefix routes and coexisting valid exact-prefix and non-invalid less-specific
alternatives, as shown in Table~\ref{tab:contexts}.

We find that some prefixes have identical collector-visible routing contexts: public BGP
collectors observe the same AS paths for their respective invalid routes and non-invalid
alternatives, indicating similar propagation patterns.  To reduce redundant measurements, we
cluster prefixes with identical collector-visible routing contexts.  From each group, we select
one representative prefix and an address within it that is not covered by any more-specific
prefix.

Using the BGP data collected at 00:00 UTC on June 30, 2026, we group 1,276 prefixes into 511 groups
and select target prefixes and addresses from 502 groups.  Note that the routes visible for a
prefix and their RPKI validation states may change over time, so the target prefixes may vary
slightly during our measurement campaign.

\subsection{IPv6 Traceroute Measurements via Open 6in4 Tunnels}

Open 6in4 tunnel endpoints have previously been used as distributed vantage points for IPv6 active
measurements~\cite{guo2025sixinfour}.  The 6in4 protocol encapsulates an IPv6 packet within an IPv4
packet~\cite{rfc4213}.  Upon receiving such a packet, an endpoint removes the outer IPv4 header and
forwards the inner packet using its IPv6 forwarding state.  Open tunnel endpoints that do not
enforce SAV on outbound traffic can therefore forward IPv6 measurement probes.

For our measurements, we leverage these tunnel endpoints to perform IPv6 traceroutes.  Specifically,
we send encapsulated ICMPv6 Echo Requests from our measurement server to a tunnel endpoint.  Each
inner IPv6 packet uses our server's IPv6 address as the source and the target address as the
destination.  We send probes with different Hop Limits in the inner IPv6 header.  The
resulting ICMPv6 responses return directly to our server, allowing us to observe the forwarding
path from the endpoint toward the target address~\cite{rfc4443}.

\begin{figure}[!t]
  \centering
  \includegraphics[width=\columnwidth]
    {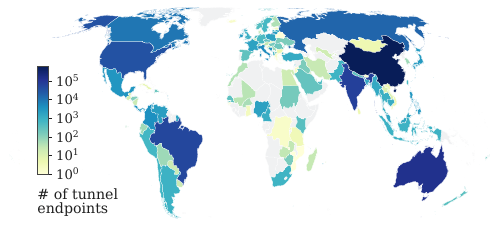}
  \caption{Geographic distribution of open 6in4 tunnel endpoints.}
  \Description{World map shaded by the number of open 6in4 tunnel endpoints
  per country on a logarithmic scale. China has the most endpoints, followed
  by Australia, India, Brazil, and the United States.}
  \label{fig:tunnel-geolocation}
\end{figure}

We identify permissive 6in4 tunnel endpoints by scanning the IPv4 address
space~\cite{guo2025sixinfour}.  Our scan finds 1,005,489 such endpoints across 2,116 IPv4 ASes.
We geolocate their IPv4 addresses using the GeoLite2 database~\cite{geolite2}.  As shown in
Figure~\ref{fig:tunnel-geolocation}, these tunnel endpoints have a broad but uneven geographic
distribution, spanning 117 countries.  Of these endpoints, 89.2\% are concentrated in China
(60.0\%), Australia (13.6\%), India (7.1\%), Brazil (5.3\%), and the United States (3.3\%).

Since the IPv4 and IPv6 interfaces of a tunnel endpoint may belong to different ASes, we infer its
IPv6 AS affiliation when possible.  First, we apply the method of Guo et
al.~\cite{guo2025sixinfour} to identify a responsive IPv6 router interface of the endpoint and
thereby obtain a direct IPv6 AS attribution.  Second, we validate an indirect attribution using
IPv6 router interfaces observed near the endpoint.  Specifically, we conduct inter-AS and intra-AS
traceroutes toward IPv6 targets outside and inside the AS associated with the endpoint's IPv4
address, respectively, and accept the attribution only when a nearby responsive IPv6 interface
also maps to that AS.  All IP-to-AS mappings are derived from RouteViews prefix-to-AS
data~\cite{routeviews}.  In total, we attribute 777,660 endpoints to 1,347 IPv6 ASes.

We conduct our traceroute measurements as follows.
\begin{enumerate}
  \item[(a)] We select representative target prefixes and addresses for measurement, as described
        in Section~\ref{sec:target-selection}.

  \item[(b)] We identify distinct vantage networks by clustering endpoints using their IPv6 ASNs
        when available, falling back to IPv4 ASNs and then IPv4 prefixes.  We randomly sample at
        most three responsive endpoints per network to reduce bias caused by uneven endpoint
        density.

  \item[(c)] From each sampled endpoint, we run IPv6 traceroutes toward all selected target
        addresses.

  \item[(d)] We conduct multiple measurement rounds throughout the campaign.  In each round, we
        reselect target prefixes and addresses using updated BGP and RPKI data, and resample
        responsive tunnel endpoints.
\end{enumerate}

\begin{figure}[!t]
  \centering
  \includegraphics[width=\columnwidth]{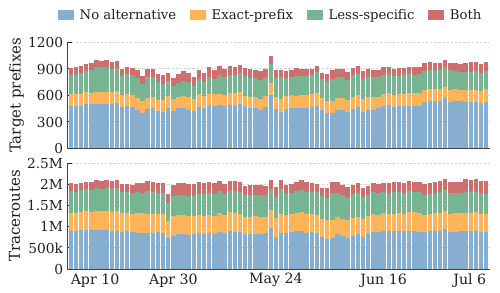}
  \caption{Measurement scale across 82 rounds, broken down by routing context.}
  \Description{Two stacked bar charts show the numbers of target prefixes
  and traceroutes per round from April to July. Each bar is split into the
  no-alternative, exact-prefix, less-specific, and both-alternatives contexts.}
  \label{fig:measurement-campaign}
\end{figure}

Between April 10 and July 6, 2026, we conducted 82 measurement rounds and collected 169.5 million
traceroutes from 1,811 vantage ASes toward 2,764 distinct target prefixes.
Figure~\ref{fig:measurement-campaign} shows the number of target prefixes and traceroutes in each
routing context for every measurement round.  For analyses based on a single round, we use the
June 30 measurement round, which contains 2,077,525 traceroutes toward 966 target prefixes
through 4,720 tunnel endpoints in 1,653 vantage ASes. During the round, we update routing
contexts using newly collected BGP data and select representative prefixes accordingly, so the
prefixes measured vary over time.

\subsection{Context-Aware Path Association}
\label{sec:association}

\textbf{Measurement challenges.} Although traceroute provides data-plane path information, determining toward which origin traffic
is forwarded is not straightforward.

\begin{figure}[!t]
  \centering
  \subfloat[]{\includegraphics[width=0.49\columnwidth]{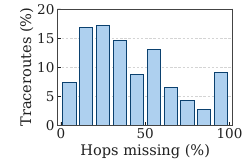}%
    \label{fig:traceroute-missing-hops}}\hfill
  \subfloat[]{\includegraphics[width=0.49\columnwidth]{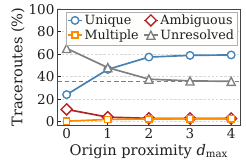}%
    \label{fig:association-scope}}
  \caption{(a) Missing-hop fractions among traces with missing hops.
    (b) Association outcomes as the origin-proximity threshold $d_{\max}$ varies.
    The dashed line marks the unresolved lower bound.}
  \Description{The left histogram bins traceroutes by their percentage of
  missing hops. The right plot shows that unique associations become more
  common and unresolved outcomes become less common as the threshold rises.
  A dashed horizontal line marks the minimum unresolved share.}
  \label{fig:path-association-diagnostics}
\end{figure}

First, traceroutes often miss IP hops along the forwarding path.
RFC~4443~\cite{rfc4443} requires IPv6 nodes to rate-limit the generation of ICMPv6 error
messages.  This mandatory rate limiting can suppress a substantial fraction of
ICMPv6 responses and has been shown to cause frequent missing hops in IPv6
traceroutes~\cite{alvarez2017ratelimiting,beholder2018ipv6}.  Explicit ICMP filtering by network
operators further exacerbates this problem~\cite{rfc4890}. Missing hop responses are also common
in our measurements. Of the 2,077,525 traceroutes collected on June 30, 1,974,243 (95.0\%) have at
least one unresponsive hop before the last responsive hop. Figure~\ref{fig:path-association-diagnostics}\subref{fig:traceroute-missing-hops}
shows how extensive these gaps are: more than half (56.1\%) of the incomplete traces are missing
at least one-third of their hops, and only 7.4\% receive responses at more than 90\% of hop positions.
Our method must therefore infer forwarding directions from partial traceroutes.

Second, even a response from within the target prefix may not uniquely identify its origin AS.
The origin can be determined from the response address only when all routes covering the probed
address share the same origin.  When different ASes originate the same prefix or overlapping
prefixes, the response address alone cannot distinguish among these origins.

\textbf{Determining forwarding directions.} We propose a context-aware path association
method that derives discriminative features from collector-visible BGP paths to determine the
forwarding direction of incomplete paths observed by traceroute.  Our analysis proceeds as
follows.

\emph{(a) Derive discriminative path features.}
        For each prefix, we derive path features from the collector-visible BGP paths in its
        routing context.  Consider a control-plane path
        \[
          C=(a_1,a_2,\ldots,a_n), \qquad a_n=o,
        \]
        where the ASes are ordered toward the origin AS $o$.  We derive three types of path
        features from $C$:

        \begin{itemize}
          \item An \emph{AS node} $a_i$ captures the presence of an individual AS on the path.
                A single hop mapped to $a_i$ is enough to match this feature, even if nearby hops
                are missing. If $a_i$ appears on BGP paths toward multiple candidate origins,
                the node alone cannot distinguish among them.

          \item An \emph{AS link} $\ell_i=(a_i,a_{i+1})$ captures a directed inter-AS adjacency
                toward the origin.  It can distinguish paths that traverse the same transit ASes
                but use different AS links.

          \item An \emph{ordered AS pair} $q_{ij}=(a_i,a_j)$ captures the relative order of two
                ASes, where $i<j$ and the two ASes need not be adjacent.  It represents the
                AS-level forwarding direction while relaxing the adjacency requirement, which
                allows us to capture directional evidence when intermediate ASes are missing from
                the observed path.
        \end{itemize}

        A feature is discriminative for origin $o$ if route collectors observe it only on BGP
        paths toward $o$. Otherwise, it is shared and cannot distinguish among those origins.

\emph{(b) Construct the observed AS path.}
        For each traceroute, we first exclude replies from addresses within the target prefix,
        since we cannot reliably identify the responding AS from such an address. We map the
        remaining reply addresses to ASes using RouteViews pfx2as data~\cite{caida_pfx2as} and use
        PeeringDB records~\cite{peeringdb} for IXP interfaces. If neither source provides a mapping,
        we fall back to GeoLite2 ASN data~\cite{geolite2}.

        To construct the observed AS path $D=(b_1,\ldots,b_m)$, we omit hops without a usable AS
        mapping and collapse consecutive occurrences of the same AS. For each resulting AS hop,
        we retain the hop numbers of its first and last replies.

\emph{(c) Match origin-proximal path features.}
        We match BGP path features against the observed AS path $D$.

        \begin{itemize}
          \item An \emph{AS node} $a_i$ matches $D$ if and only if
                $\exists\, r,\ 1\le r\le m:\ b_r=a_i$.

          \item An \emph{AS link} $\ell_i=(a_i,a_{i+1})$ matches if $a_i$ and $a_{i+1}$ occur
                consecutively in $D$ in that order, and the last reply from $a_i$ and the first
                reply from $a_{i+1}$ have consecutive hop numbers in the original traceroute.

          \item An \emph{ordered AS pair} $q_{ij}=(a_i,a_j)$ matches $D$ if and only if
                $\exists\, r,s,\ 1\le r<s\le m:\ (b_r,b_s)=(a_i,a_j)$.
        \end{itemize}

        When the observed AS path matches a feature discriminative for origin $o$, the traceroute
        reveals interdomain forwarding behavior partially consistent with a BGP path toward $o$.
        However, this does not establish that traffic follows the remaining BGP-path suffix to
        $o$.  We therefore characterize the feature's \emph{origin proximity} by its AS-hop
        distance to $o$ and denote this distance by $k$.  A smaller $k$ requires less
        extrapolation to infer forwarding direction.

        For a control-plane path $C=(a_1,\ldots,a_n)$ with origin $o=a_n$, the origin distances of
        AS node $a_i$, AS link $\ell_i=(a_i,a_{i+1})$, and ordered AS pair
        $q_{ij}=(a_i,a_j)$ are $n-i$, $n-i-1$, and $n-j$, respectively.  For the latter two
        features, the distance is measured from the AS closer to $o$.  When a feature appears on
        multiple BGP paths toward the same origin, we compare each path with $D$ using the longest
        common subsequence (LCS) and use the largest origin distance among the paths with the
        maximum LCS length.  For each origin, we use the minimum origin distance among all matched
        discriminative features.

        To limit path extrapolation toward the origin, we introduce an origin-proximity threshold
        $d_{\max}$.  We associate $D$ with $o$ only when a matched discriminative feature lies
        within $d_{\max}$ AS hops of $o$.

\emph{(d) Infer forwarding directions.}
        Finally, we classify each traceroute into one of four association outcomes based on its
        observed AS path $D$:
        \begin{itemize}
          \item \emph{Unique association}: $D$ matches discriminative features within $d_{\max}$
                toward either an invalid or an alternative origin.
          \item \emph{Multiple association}: $D$ matches discriminative features within $d_{\max}$
                toward both invalid and alternative origins.
          \item \emph{Ambiguous association}: $D$ matches no discriminative feature within
                $d_{\max}$ but matches at least one shared feature.
          \item \emph{Unresolved}: $D$ matches neither a discriminative feature within $d_{\max}$
                nor a shared feature.
        \end{itemize}

        A unique association identifies forwarding toward invalid or alternative
        origins. A multiple association shows
        path-level indications of forwarding toward both invalid and alternative
        origins, without identifying a unique forwarding direction.
        Ambiguous and unresolved associations do not determine a forwarding direction.

For analyses of invalid-origin forwarding, we use path features that distinguish
invalid routes from non-invalid alternatives without requiring a unique origin
ASN. A feature shared by multiple invalid origins remains discriminative for the
invalid direction if it does not occur on alternative routes; the same principle
applies to the alternative direction. When identifying individual origins, we
require features specific to the corresponding origin ASN.

\textbf{Origin-proximity threshold.} Selecting $d_{\max}$ involves a trade-off between association coverage and the proximity of
matched features to the origin.  A smaller threshold restricts associations to features closer
to the origin and reduces path extrapolation, but leaves more traceroutes unresolved or
ambiguous.  A larger threshold admits features farther upstream, which may identify additional
forwarding directions but may also produce matches toward both invalid and alternative origins.

We evaluate $d_{\max}\in\{0,\ldots,4\}$ to characterize this trade-off.  As shown in
Figure~\ref{fig:path-association-diagnostics}\subref{fig:association-scope}, increasing $d_{\max}$ from zero to two raises unique
associations from 23.90\% to 57.28\% and reduces unresolved traces from 65.00\% to 37.66\%.
Multiple associations increase only from 0.31\% to 2.24\%, indicating that most additional
matches identify a unique forwarding direction.

Increasing $d_{\max}$ further to four yields only modest gains: unique associations reach
59.18\%, while unresolved traces decline to 35.71\%. The horizontal line marks the 35.63\%
of traceroutes that share no AS with any BGP path in their routing context and therefore
remain unresolved regardless of the threshold. The unresolved fraction at $d_{\max}=2$
is already close to this lower bound. We therefore adopt $d_{\max}=2$, retaining most
of the association coverage while limiting path extrapolation.

\section{Characterizing Data-Plane Forwarding}
\label{sec:results}

In this section, we first characterize forwarding directions across routing contexts and
investigate unresolved traceroutes. We then assess the consistency of forwarding
observations within vantage networks and over time.

\subsection{Where Is Traffic Forwarded?}
\label{sec:origin-direction}

\begin{figure}[!t]
  \centering
  \includegraphics[width=\columnwidth]{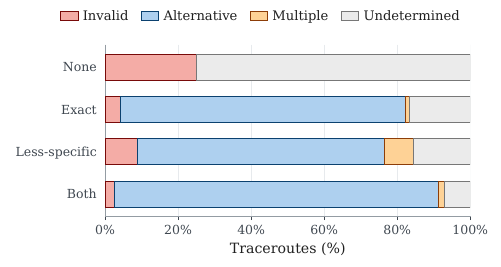}
  \caption{Forwarding directions by routing context.}
  \Description{Four horizontal stacked bars show the shares of invalid,
  alternative, multiple, and undetermined forwarding. Most traceroutes are
  undetermined when no alternative exists. Alternative forwarding accounts
  for the largest share in each of the other three routing contexts.}
  \label{fig:origin-forwarding}
\end{figure}

Figure~\ref{fig:origin-forwarding} summarizes forwarding directions for all 2,077,525 traceroutes
across routing contexts. We group ambiguous and unresolved associations as
\emph{undetermined} and refer to forwarding identified only toward invalid origins as
\emph{invalid-origin forwarding}.

\textbf{A quarter of traceroutes in the no-alternative context are forwarded toward invalid
origins.}
Among 899,957 traceroutes toward 534 prefixes in the no-alternative context, 25.1\% are forwarded
toward invalid origins; the remaining 74.9\% are unresolved.
Section~\ref{sec:unresolved-diagnostics} examines whether missing responses or unavailable invalid
routes can account for these unresolved traceroutes.

\textbf{Non-invalid alternatives dominate forwarding when available.}
Alternative-origin forwarding accounts for 77.9\% of all traceroutes in the exact-prefix
context, 67.8\% in the less-specific context, and 88.8\% in the both-alternatives context.
Among traceroutes with a unique forwarding direction, 88.6--97.2\% are directed toward
alternative origins across these three contexts.

\textbf{Some traceroutes exhibit forwarding along both invalid and alternative routes.}
Multiple associations account for 2.2\% of all traceroutes. They are most
frequent in the less-specific context (7.8\%), compared with 1.2\% in the
exact-prefix context and 1.4\% in the both-alternatives context.
With less-specific alternatives, traffic often follows a valid covering
route before being diverted onto the more-specific invalid route. With
exact-prefix alternatives, traffic more often switches from invalid to valid
routes. We observe that some invalid origins act as transit ASes and forward
the traffic onward toward the valid origin. When both types of alternatives
exist, we observe traffic redirected from valid to invalid routes as well as
from invalid to valid routes. Appendix~\ref{app:multiple-forwarding} presents
the path analysis and representative cases for each context.

\subsection{Why Do Traceroutes Remain Unresolved?}
\label{sec:unresolved-diagnostics}

In the no-alternative context, 674,044 traceroutes (74.9\%) remain unresolved.  We examine whether
this prevalence arises from missing hop responses, which leave the observed AS paths without
sufficient discriminative path features, or from invalid-route unavailability, which causes the
observed forwarding paths to terminate prematurely.

\begin{figure}[!t]
  \centering
  \includegraphics[width=3.33in]{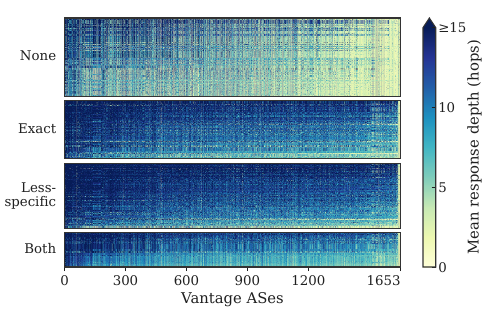}
  \caption{Mean traceroute response depth for prefix--AS pairs, grouped by routing
  context.}
  \label{fig:response-depth-context}
  \Description{An AS-by-prefix heatmap groups rows into no-alternative,
  exact-prefix, less-specific, and both-alternatives contexts, with one shared linear color scale.
  AS columns and prefixes within each group are ordered from deeper to shallower mean responses.
  The horizontal axis spans 0 to 1,653, marking column boundaries in the sorted AS list.
  The no-alternative block is generally lighter, indicating shallower responses.
  Gray cells indicate pairs without usable response records.}
\end{figure}

\textbf{No-alternative traceroutes are more likely to terminate prematurely.}
We define the \emph{response depth} of a traceroute as its last responsive hop.
For Figure~\ref{fig:response-depth-context}, we retain prefixes with response records from
at least 80\% of vantage ASes.
As shown in Figure~\ref{fig:response-depth-context}, traceroutes toward prefixes
without alternatives generally terminate prematurely, with a median response
depth of 5 hops, compared with 10--13 for prefixes with alternatives.
Among unresolved traceroutes, 26.1\% of those without alternatives end
with an ICMPv6 \emph{No Route to Destination} message~\cite{rfc4443},
compared with 6.1--13.7\% of those with alternatives.

\begin{figure}[!t]
  \centering
  \includegraphics[width=.91\linewidth]{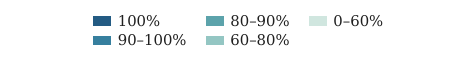}\par
  \includegraphics[width=\linewidth]{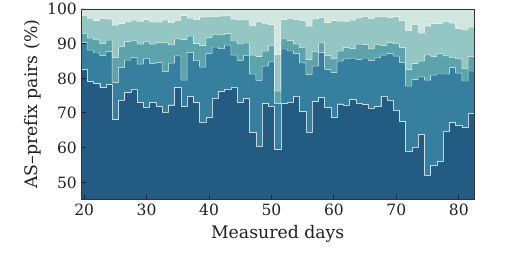}
  \caption{Temporal consistency by measurement duration.}
  \Description{Stacked bands show how AS--prefix pairs are distributed across
  five consistency bins for each number of measured days. At most durations,
  the largest band represents pairs whose classification is the same on
  every measured day.}
  \label{fig:prefix-vantage-stability}
\end{figure}

\textbf{Forwarding resumes when routes become valid.}
To distinguish early forwarding termination from missing ICMPv6 responses,
we remeasure 25 prefixes after their routes change from RPKI-invalid to
RPKI-valid. We repeat 55,813 traceroutes from the same tunnel endpoints to
the same destination addresses. Of the 35,864 previously
unresolved traceroutes, 90.8\% receive responses from farther hops, with the median
response depth increasing from 4 to 12. By contrast, the median response
depth of traceroutes already directed toward the origin does not increase.
Among the previously unresolved traceroutes, the percentage receiving ICMPv6
\emph{Destination Unreachable} messages also falls from 29.2\% to 8.7\%.
These observations suggest that missing ICMPv6 responses were not the primary
cause of the unresolved paths.

We further examine whether the recovered paths show forwarding toward the
origin. Of the previously unresolved traceroutes, 32,748 (91.3\%) resume
forwarding toward the origin, while 96.8\% of those previously directed
toward the invalid origin continue toward the same origin. The recovery
indicates that route unavailability was the main cause of the earlier
unresolved traceroutes, likely because route filtering had limited the
propagation of the invalid routes.
\subsection{How Consistent Is Forwarding Behavior?}
\label{sec:persistence}

We evaluate forwarding consistency by asking two questions: whether
traceroutes from the same AS to a given prefix agree on the forwarding
direction, and whether forwarding observations for the same AS--prefix pair
remain stable over time.

\textbf{Forwarding directions for a given prefix are consistent within vantage networks.}
We evaluate 580,319 AS--prefix pairs with multiple traceroute observations,
accounting for 89.4\% of all traceroutes. Of these pairs, 99.7\% show no conflict
between determined forwarding directions. Specifically, among the 85,081 pairs
with at least one invalid-origin direction, 97.9\% show no alternative-origin
forwarding. These results support characterizing AS-level forwarding behavior
toward individual prefixes from our observations.

\textbf{Forwarding observations are consistent over time.}
We examine 921,413 AS--prefix pairs measured on at least 20 days with no change
in routing context. For each AS--prefix pair, we measure consistency by how
often its most common daily classification recurs across measurement days.
As Figure~\ref{fig:prefix-vantage-stability} shows, 88.4\% of pairs receive
the same classification on at least 80\% of measured days, and 70.8\%
receive it on every measured day.
Of the 247,963 AS--prefix pairs with observed invalid-origin forwarding,
86.0\% show it on multiple measurement days.

\begin{figure}[!t]
  \centering
  \includegraphics[width=\linewidth]{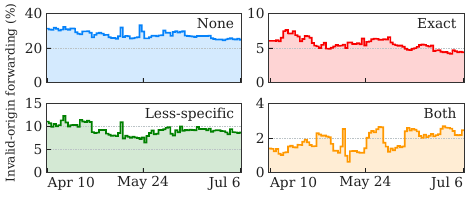}
  \caption{Invalid-origin forwarding by routing context across 82 rounds.}
  \Description{Four time series show the share of traceroutes forwarded toward
  an invalid origin in each routing context. The no-alternative series stays
  above the other three across the measurement campaign. The panels use
  different vertical scales.}
  \label{fig:forwarding-prevalence}
\end{figure}

\begin{figure}[!t]
  \centering
  \includegraphics[width=\linewidth]{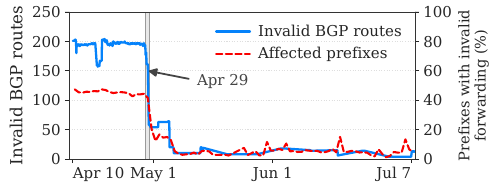}
  \caption{Invalid routes and invalid-origin forwarding through AS37468.}
  \label{fig:as37468-timeline}
  \Description{A single plot overlays control-plane and data-plane time
  series for 438 IPv6 prefixes from April 10 through July 7 using separate
  labeled vertical axes. A solid blue line counts invalid BGP routes
  visible through AS37468; a dashed red line reports the daily percentage
  of prefixes observed through the AS with at least one invalid-origin
  forwarding traceroute. Both decline around April 29.}
\end{figure}

\textbf{Invalid-origin forwarding declined modestly overall.}
As shown in Figure~\ref{fig:forwarding-prevalence}, invalid-origin forwarding
generally declined across all routing contexts except the both-alternatives
context, where it fluctuated below 3\%. Across all contexts, its share fell
from 17.9\% of traceroutes in the first
7 measurement rounds to 14.0\% in the last 7 rounds. Stronger ROV enforcement
at some transit ASes likely contributed to the declines by limiting the
propagation and use of invalid routes.

For example, AS37468 (Angola Cables), a global IP transit provider, appears
to have strengthened its IPv6 ROV enforcement in late April.
Figure~\ref{fig:as37468-timeline} tracks control- and data-plane observations
throughout the measurement period for 438 IPv6 prefixes whose routing
contexts remained unchanged from April~22 to May~6. Although their invalid
origins continued to announce these prefixes throughout this interval,
invalid routes visible through
AS37468 dropped sharply on April~29, from about 200 before the change to
about 60 by 16:00 UTC that day. After May~6, only 4--20 remained visible
through the AS.

Data-plane measurements confirm a corresponding shift in forwarding
behavior. The data-plane curve plots the daily share of prefixes observed
through AS37468 with at least one traceroute forwarded toward an invalid
origin. Before the April~29 routing change, 45.4\% of prefixes exhibited
such forwarding on average; this share dropped to 20.1\% the next day.
After May~6, invalid-origin forwarding remained observable for only 5.0\%
of prefixes on average.

\section{Anatomy of Invalid-Origin Forwarding}
\label{sec:invalid-origin-forwarding}

We first characterize the prevalence of invalid-origin forwarding across
prefixes and ASes, then examine its relationship with the invalid origin,
alternative-route availability, and ROV protection. We next
combine operational attribution with BGP and traceroute observations to explain
why some invalid routes have widespread forwarding impact while others remain
limited. Finally, we examine why some prefixes have multiple invalid origins
and whether traffic is forwarded toward each origin.

\subsection{How Widespread Is Invalid-Origin Forwarding?}
\label{sec:heterogeneity}

\begin{figure}[!t]
  \centering
  \includegraphics[width=3.33in]{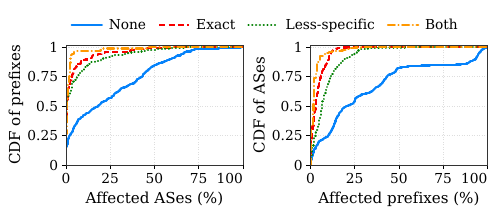}
  \caption{Prevalence of invalid-origin forwarding across prefixes (left) and ASes (right).}
  \label{fig:origin-heterogeneity}
  \label{fig:origin-heterogeneity-prefix}
  \label{fig:origin-heterogeneity-vantage}
  \Description{Two empirical cumulative distributions compare forwarding coverage across
  measured prefixes and ASes. A shared legend above both panels distinguishes
  routing contexts by color and line style. Coverage is generally higher without alternatives.}
\end{figure}

Invalid-origin forwarding accounts for 24.7\% of traceroutes with a unique
forwarding direction. We examine its prevalence across measured prefixes and ASes.
We count an AS--prefix pair as affected if at least one traceroute is classified
as forwarding toward an invalid origin and none is classified as forwarding
toward an alternative origin. For each routing context,
Figure~\ref{fig:origin-heterogeneity} includes prefixes measured from at least
20 ASes and ASes measured against at least 20 prefixes.

\textbf{Invalid-origin forwarding spans most measured prefixes.}
We observe it from at least one AS for 747 of the 956 prefixes (78.1\%).
As Figure~\ref{fig:origin-heterogeneity-prefix} (left) shows, the proportion of
ASes forwarding toward invalid origins varies substantially across prefixes
and is generally lower when alternatives exist. For 47.3\% of prefixes without
alternatives, more than 20\% of ASes forward traffic toward invalid origins.
For 89.1\% of prefixes with exact alternatives and 81.2\% with less-specific
alternatives, at most 10\% of ASes forward traffic toward invalid origins.
With both types of alternatives, such forwarding is observed from at most
5\% of ASes for 96.7\% of prefixes.

\textbf{Invalid-origin forwarding is observed from nearly all ASes.}
Across all routing contexts, we observe it from 1,643 of the 1,653 ASes
(99.4\%). Figure~\ref{fig:origin-heterogeneity-vantage} (right) shows substantial
variation in how broadly ASes forward toward invalid origins across their
measured prefixes, particularly without alternatives. Without alternatives,
12.2\% of ASes forward traffic toward invalid origins for more than 90\% of their
measured prefixes. With alternatives, however, invalid-origin forwarding
involves at most 20\% of measured prefixes for more than 90\% of ASes in each
context.

\subsection{What Shapes Forwarding Behavior?}
\label{sec:forwarding-factors}
\label{sec:same-origin-forwarding}
\label{sec:origin-network-variation}

We examine how invalid-origin forwarding varies with the invalid origin,
alternative-route availability, and ROV protection in measured networks.

\textbf{Invalid routes for same-origin prefixes have more similar data-plane
impacts.} We select 1,621 same-origin prefix pairs, covering 418 prefixes
announced by 72 origins. The two prefixes in each pair share a routing context
and are measured from at least 50 common ASes. Within these common ASes, let
$A_p$ and $A_q$ denote those with invalid-origin forwarding for prefixes $p$
and $q$, respectively. We measure their overlap using the Jaccard index:
\[
J(p,q)=\frac{|A_p\cap A_q|}{|A_p\cup A_q|}.
\]

We then assess whether prefixes that share an origin tend to have greater
overlap. For each same-origin pair, we construct a matched control set by
randomly selecting up to 50 different-origin pairs in the same routing context,
with a similar number of ASes from which both prefixes were measured and
similar affected-AS proportions for the corresponding prefixes. The control
sets contain 47.6 pairs on average. Within each set, we compute the mean
Jaccard overlap and the share of pairs with at least 80\% overlap. We then
average each statistic across the 1,621 control sets, weighting them equally
to obtain the matched baseline.

The mean overlap is 45.2\% for same-origin pairs and 15.7\% for the matched
baseline. The share of pairs with at least 80\% overlap is 34.5\% for
same-origin pairs and 8.2\% for the matched baseline.
These results show that invalid routes for same-origin prefixes tend to affect
similar ASes. Shared interconnections and export policies may explain this
similarity, although prefix-specific policies can produce different BGP
paths~\cite{wu2025policyatoms}.

\textbf{Same-origin comparisons confirm less widespread invalid-origin
forwarding with alternatives.} We further examine whether the smaller
forwarding scope associated with alternatives also holds among prefixes
announced by the same invalid origin. We identify 15 invalid origins that each
announce prefixes both with and without non-invalid alternatives, covering
170 prefixes. Across these origins, invalid-origin forwarding is observed
from an average of 20.1\% of ASes for prefixes without alternatives and 4.7\%
for those with alternatives. These results reinforce the role of alternative
routes in limiting invalid-origin forwarding, even among prefixes announced
by the same origin.

\begin{figure}[!t]
  \centering
  \includegraphics[width=\columnwidth]{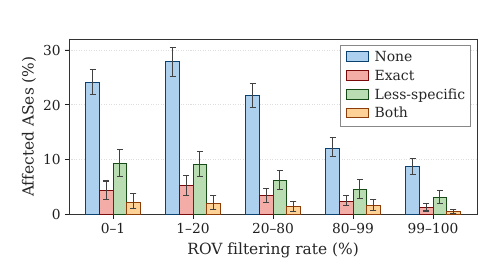}
  \caption{Invalid-origin forwarding by APNIC IPv6 ROV filtering rate.
  Error bars show 95\% bootstrap confidence intervals (3,000 prefix resamples).}
  \Description{Grouped bars compare four routing contexts across five ROV
  filtering-rate bins. The share of ASes forwarding toward invalid origins
  is highest without alternatives and generally declines at higher filtering
  rates in all four contexts.}
  \label{fig:apnic-invalid-forwarding}
\end{figure}

\textbf{ASes with stronger ROV protection are less likely to forward toward
invalid origins.} We next assess the role of ROV in invalid-origin
forwarding. We use the IPv6 ROV filtering rates reported by APNIC to
characterize the protection provided by an AS or its upstream
networks~\cite{apnic2021rov}. We divide ASes into five groups
based on these rates, as shown in Figure~\ref{fig:apnic-invalid-forwarding}.
Within each routing context, we restrict our analysis to prefixes observed
across all five groups so that every group is evaluated on the exact same
set of prefixes. For each prefix, we calculate the share of measured ASes
in each group that forward toward invalid origins, and then average these
shares across prefixes with equal weight.

Figure~\ref{fig:apnic-invalid-forwarding} shows generally less invalid-origin
forwarding as ROV protection strengthens across all four routing contexts.
Within each group, ASes are also less likely to forward toward invalid
origins for prefixes with alternatives, consistent with our earlier
findings. However, even among ASes with filtering rates above 99\%,
invalid-origin forwarding remains observable from 8.7\% of ASes for
prefixes without alternatives and 0.5--3.1\% for those with alternatives.
This suggests that high APNIC filtering rates do not imply complete ROV
protection.

\subsection{Why Do Some Invalid Prefixes Have Widespread Impact?}
\label{sec:operational-explanations}
\label{sec:widespread-invalid-forwarding}

We observe invalid-origin forwarding from more than half of measured ASes
for 85 prefixes: 80 without alternatives and 5 with alternatives. We study
these prefixes to understand why this behavior is widespread.

\textbf{Most prefixes with widespread forwarding have identifiable operational
explanations.} We identify operational explanations for prefixes without
alternatives through attribution
analysis~\cite{li2026demystifying} and IRR records
(Appendix~\ref{app:attribution}).
Specifically, 39 exceed the
maxLength permitted by same-origin ROAs; 28 have BGP origins operated by the same organizations
as their ROA-authorized ASes; and another 3 have BGP origins with a direct
provider--customer relationship with a ROA-authorized AS. Another 5 prefixes are announced by AS13527
(GOOFL-TESTBED) from address space explicitly registered for experimental RPKI
measurement. Its provider AS20473 (Vultr) maintains covering IRR route objects
identifying them as customer routes~\cite{ripestat_gooflare_irr}.
The remaining 5 prefixes include Cloudflare's ROV test prefix, announced
from its globally distributed PoPs~\cite{poinsignon2020bgpsafe}, and 4 other
prefixes that propagate through major transit ASes such as AS6762 (Telecom
Italia Sparkle) and AS60068 (CDN77)~\cite{routeviews}.

When no alternative route is available, operators may be more inclined to
accept and propagate invalid routes with identifiable operational explanations
to preserve connectivity~\cite{li2023rovista,li2026demystifying,schulmann2025conflicts}.
To investigate this possibility, we compare attributed and unexplained prefixes
with and without alternatives (Appendix~\ref{app:attribution}). The difference
is pronounced only when no alternative is available: on average, 30.1\% of
measured ASes forward toward invalid origins for attributed prefixes, versus
9.7\% for unexplained prefixes. With alternatives, the corresponding
percentages are nearly identical at 4.5\% and 4.7\% respectively. This contrast
suggests that operators may deliberately relax ROV enforcement to preserve
connectivity when no alternative route is available.

We next investigate why invalid-origin forwarding remains widespread for
the 5 prefixes with alternatives.

\begin{figure}[!t]
  \centering
  \includegraphics[width=0.90\linewidth,clip]{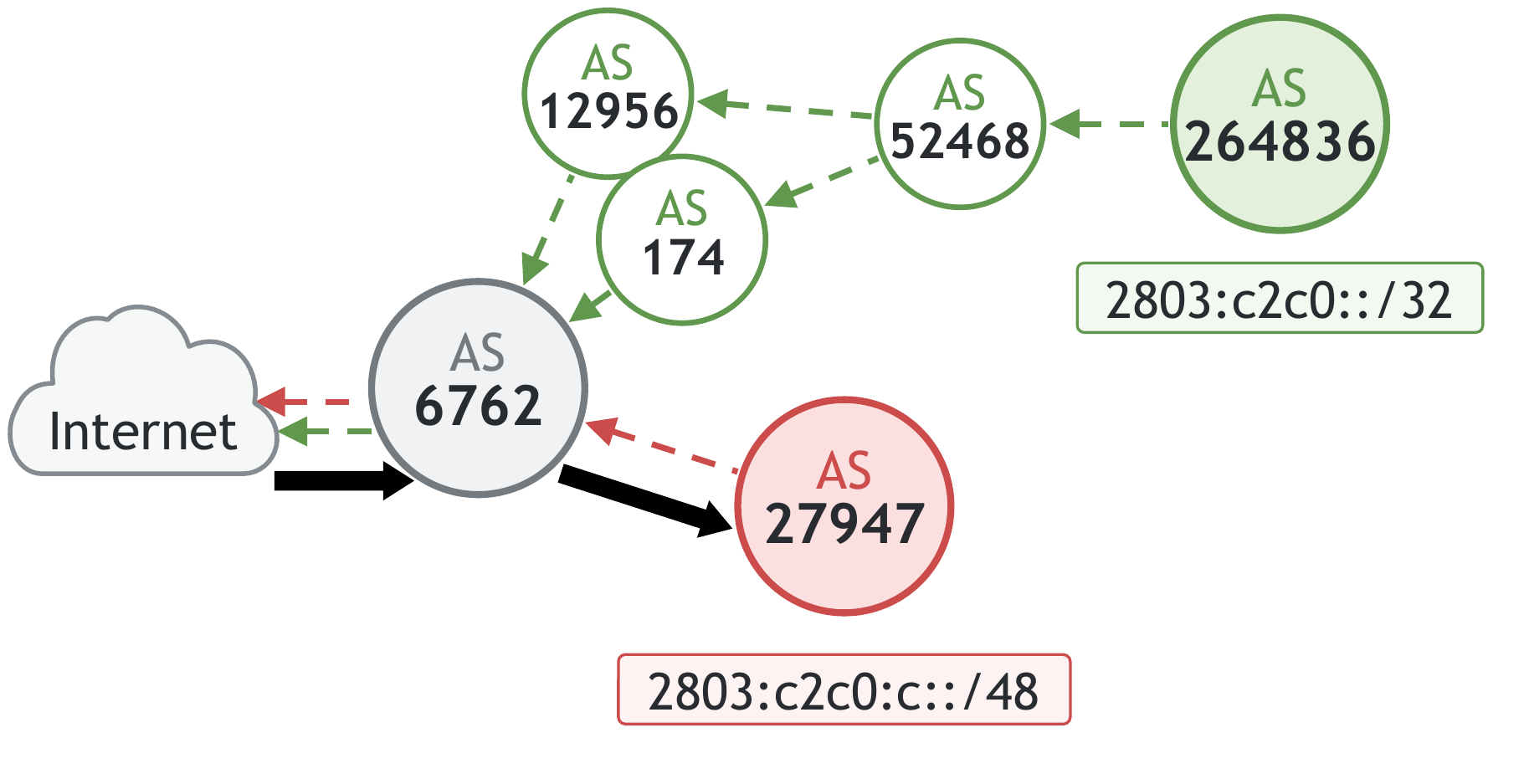}
  \caption{AS6762 diverts traffic from the valid covering route onto the
  invalid \texttt{2803:c2c0:c::/48} route.}
  \Description{A single-column network diagram shows the Internet and AS6762
  on the left. Green dashed arrows show propagation of the valid covering
  route from AS264836 through AS52468 and AS12956 or AS174. Red dashed arrows
  show propagation of the invalid /48 from AS27947 through AS6762. Black
  solid arrows show traffic forwarded from the Internet through AS6762 to AS27947.}
  \label{fig:less-specific-diversion}
\end{figure}

\textbf{More-specific invalid routes divert traffic from valid covering
routes.} For three prefixes, the ROA-authorized AS originates a valid
covering route, while its provider originates a more-specific invalid route.
Figure~\ref{fig:less-specific-diversion} illustrates one such case:
AS264836 (APROSVA) originates the valid
\texttt{2803:\allowbreak c2c0::/32}, while its provider AS27947 (Telconet
Ecuador) originates the invalid \texttt{2803:\allowbreak c2c0:c::/48}.
AS6762 learns both routes through different next hops. In
RouteViews~\cite{routeviews}, the valid covering /32 is observed
across 259 ASes, whereas the invalid /48 is observed across only 55.
Nevertheless, invalid-origin forwarding dominates across all three
prefixes: 53.3--64.7\% of traceroutes for each prefix are directed toward
the invalid origin, versus 1.7--12.0\% toward the valid origin.

A further 27.0--38.4\% of traceroutes per prefix show evidence of forwarding
along both valid and invalid routes. We verify that 97.0--98.8\% of these
traceroutes initially follow the covering route before being redirected
toward the invalid origin. AS6762 is a key transit in this diversion:
longest-prefix matching directs traffic arriving along the covering route
to AS27947 using the more-specific invalid route. Such downstream diversion
can cause collateral damage under partial ROV deployment even when upstream
networks filter invalid routes~\cite{gilad2016deployment,li2023rovista,li2025improv}.
Appendix~\ref{app:multiple-less-specific} provides the supporting path analysis.

\begin{figure}[!t]
  \centering
  \includegraphics[width=0.90\linewidth,clip]{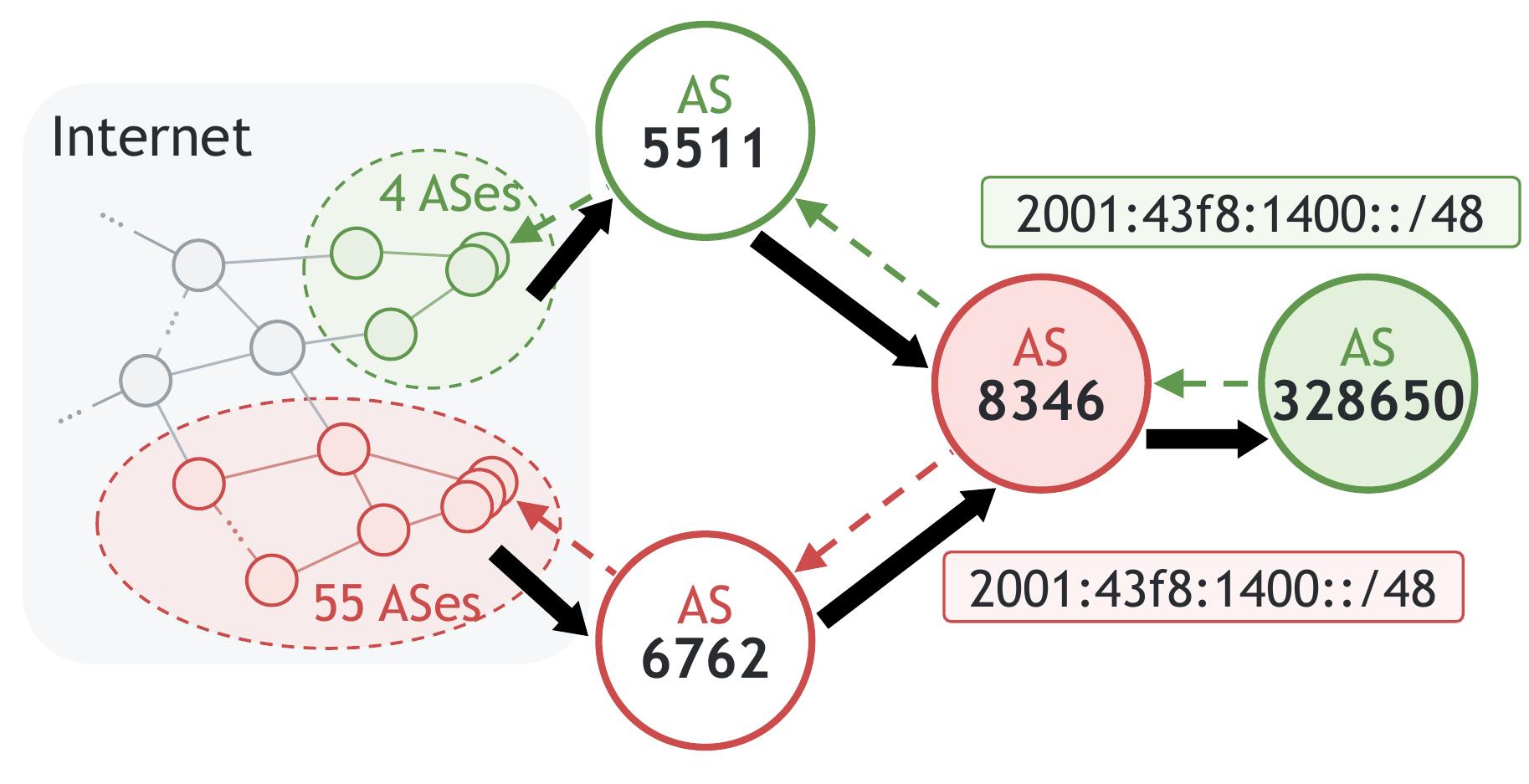}
  \caption{The invalid route for \texttt{2001:\allowbreak43f8:1400::/48}
  propagates much more widely than the valid route.}
  \Description{A single-column network diagram shows representative Internet
  ASes in green and red shaded regions. The valid route originates at
  AS328650 and propagates through AS8346 and AS5511. The invalid route
  originates at AS8346 and propagates through AS6762. Black arrows show
  the corresponding traffic directions toward the two origins.}
  \label{fig:exact-prefix-transit-asymmetry}
\end{figure}

\textbf{Asymmetric transit support favors an invalid exact-prefix route.} For
\texttt{2001:\allowbreak43f8:1400::/48}, valid and invalid exact-prefix routes
coexist but receive sharply different transit support. As illustrated in
Figure~\ref{fig:exact-prefix-transit-asymmetry}, AS328650 originates the valid
route, which its provider AS8346 advertises to AS5511. AS8346 also originates
an invalid route for the same prefix and advertises it to AS6762. In
RouteViews~\cite{routeviews}, the valid route is observed across only 4 ASes,
versus 55 for the invalid route. Because AS8346 appears on paths toward both
origins, we distinguish forwarding along the two routes using their
respective transit branches through AS5511 and AS6762. In the data plane,
61.0\% of traceroutes are directed toward the invalid origin AS8346; 99.8\%
of these traverse AS6762. Only one traceroute is directed toward the valid
origin AS328650, through AS5511.

To understand why the valid alternative attracts so little traffic, we
jointly analyze historical BGP and traceroute observations. We identify a
sharp contraction in its propagation through AS5511 between June 16 and 17,
when the number of distinct observed AS paths fell from 223 to 2 while the
invalid route expanded through AS6762. This contraction was specific to the
prefix: other observed valid routes through AS5511 and AS8346 remained
stable. Our traceroutes reveal a corresponding shift in forwarding: the
percentage of measured ASes forwarding toward the valid origin fell from
32.1\% to 2.6\%, while that forwarding toward the invalid origin rose from
39.3\% to 59.5\%. Appendix~\ref{app:exact-prefix-evolution} presents this
case in more detail.

\textbf{Extensive interconnection sustains invalid-route propagation.} For
\texttt{2a05:dfc7:30::/48}, AS36351 (IBM Cloud/SoftLayer)
originates the invalid /48, while AS58212 (Dataforest) announces the
covering \texttt{2a05:}\allowbreak\texttt{dfc7::/32}. Despite this alternative,
53.4\% of traceroutes are directed toward the invalid origin AS36351, while
42.4\% are directed toward AS58212. The covering /32 propagates through
major transit providers such as AS6939 (Hurricane Electric) and AS2914
(NTT). In contrast, the invalid /48 is not observed to propagate through
these providers, which may be due to their ROV filtering
policies~\cite{he_routing_policy,ntt_rpki_policy}. Nevertheless, AS36351's
extensive interconnection provides other propagation paths. The network has
a documented presence at 54 IXPs and 40 facilities~\cite{peeringdb}, and
RouteViews observations show that the invalid route reaches at least 98
ASes, including 61 direct neighbors of AS36351~\cite{routeviews}. These
direct connections allow the route to propagate without traversing those
transit providers.

\subsection{Why Do Others Have Limited Impact?}
\label{sec:limited-invalid-forwarding}

We next focus on 596 prefixes (62.3\% of the 956 prefixes) for which
invalid-origin forwarding is observed from no more than 10\% of measured
ASes. We examine how routing policies and changes in ROA authorization
limit the propagation and data-plane impact of invalid routes.

\textbf{Filtering and export policies restrict invalid-route propagation.}
We analyze the RPKI-invalid route that AS398549 (Zero Attack Vector)
originates for \texttt{2606:\allowbreak a300:1005::/48}. Its transit
provider, AS35847 (Ezee Fiber), announces a globally visible covering
route for \texttt{2606:\allowbreak a300:1000::/40}. AS35847's major
upstreams, AS1299 (Arelion) and AS2914 (NTT), both propagate the four
other RPKI-valid /48s from AS398549, yet neither propagates the invalid
/48~\cite{routeviews,ripe_ris}. This is consistent with their published
ROV policies~\cite{arelion_rpki_policy,ntt_rpki_policy}.

We observe the invalid /48 only along the AS path
\texttt{AS49544 AS35847 AS398549}. The route's BGP communities indicate
that AS49544 learns it from AS35847 via public
peering~\cite{i3d_bgp_communities}. Because the route is peer-learned,
AS49544's export policy restricts its advertisement strictly to its
customers~\cite{ripe_as49544_policy}. Consequently, the covering alternative
dominates data-plane forwarding, with 98.3\% of traceroutes directed
toward AS35847.

\textbf{Loss of ROA authorization limits route propagation.} During our
measurement campaign, we identify 73 prefixes from 8 origins that lose ROA
authorization while their BGP origins and direct upstreams remain unchanged.
For 66 prefixes, authorization shifts to other origins; for the remaining 7,
the original authorizations are withdrawn, leaving only AS0 ROAs. Across
these prefixes, observed propagation contracts from a median of 256 ASes
while valid to 17 after becoming invalid (as observed on June 30).

Two cases demonstrate this dynamic. For
\texttt{2a06:\allowbreak1184:100::/48}, authorization shifts from AS154049
(EXA TECH) to AS60781 (LeaseWeb Netherlands). The address space is
registered to RapidSeedbox, an IP leasing
provider~\cite{ripe_rapidseedbox_allocation,rapidseedbox_ip_leasing} that lists
AS60781 as the origin in the prefix's IRR route
object~\cite{ripe_rapidseedbox_route6}.
AS154049 continues announcing the prefix through AS21859 (Zenlayer), but
its observed propagation contracts from 259 to 7 ASes. For
\texttt{2a13:\allowbreak be40::/29}, AS401776 (IDC Cube) continues announcing
the prefix through AS62610 (Zenlayer) after its authorization is withdrawn,
leaving only AS0 ROAs; propagation contracts from 243 to 14 ASes. Both routes
have limited data-plane impact: invalid-origin forwarding accounts for only
0.4\% and 0.2\% of traceroutes for the two prefixes, respectively.


\subsection{Which Invalid Origin Attracts the Traffic?}
\label{sec:multiple-invalid-origins}

We identify 23 prefixes announced by two invalid origins in our measurements.
For 11 prefixes, separate traceroutes are directed toward each invalid origin.
For another 10, invalid-origin forwarding is observed toward only one of the
two origins. For the remaining 2, most traceroutes follow valid alternatives
or remain unresolved. To interpret these observations, we contextualize them
with operational background and historical BGP and ROA records.

\textbf{Both invalid origins remain in use, but forwarding consistently favors one.}
For 9 prefixes, ROA authorization shifted from AS154049 (EXA TECH) to AS60781
(Leaseweb Netherlands). AS154049 continues announcing them, while AS30633
(Leaseweb USA) also originates them through AS16265 (Leaseweb Network).
AS30633, AS16265, and the authorized AS60781 all belong to Leaseweb. The two
origins likely become invalid for distinct reasons: an intra-organizational
ROA omission for AS30633 and unsynchronized ROA and BGP updates for AS154049.

Using origin-specific path features, we distinguish forwarding directions
even without directly observing the origin AS. Across the 9 prefixes, we
directly observe AS30633 in 78.7\% of traceroutes identified as forwarding
toward it. In contrast, we identify 211 traceroutes as forwarding toward
AS154049 using origin-specific path features, without directly observing
the origin AS. Taking \texttt{2a0e:\allowbreak1a80:2541::/48} as an example,
we observe traceroutes reaching AS30633 directly, such as
$\mathrm{AS32098}\!\to\!\mathrm{AS16265}\!\to\!\mathrm{AS30633}$.
For AS154049, we use origin-specific path features such as the AS138997 node
and the AS8966--AS138997 link. These features identify forwarding toward
AS154049 in the traceroute
$\mathrm{AS14593}\!\to\!\mathrm{AS4800}\!\to\!\mathrm{AS8966}\!\to\!\mathrm{AS138997}$.

Finally, we observe forwarding toward both invalid origins for all 9
prefixes, with
more traceroutes directed toward AS30633 in every case. Of all traceroutes
measured for these prefixes, 5.6\% are directed toward AS30633 and 2.0\%
toward AS154049; the remaining traceroutes are unresolved.
We discuss another case involving regional origins of the same operator in
Appendix~\ref{app:regional-origin-forwarding}.

\section{Evaluation}
\label{sec:association-diagnostics}

In this section, we evaluate the effectiveness of path features, validate inference accuracy
and recovery under response loss, and assess robustness to limited
control-plane visibility and temporal offsets.

\subsection{Effectiveness of Path Features}
\label{sec:path-features}

\begin{figure}[!t]
    \centering
    \includegraphics[width=0.49\linewidth]{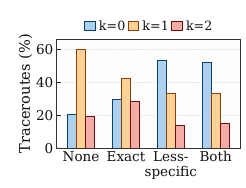}\hfill%
    \includegraphics[width=0.49\linewidth]{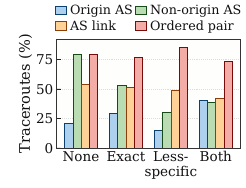}
    \caption{Origin proximity (left) and types (right) of selected path features by routing context.}
    \Description{Two grouped bar charts compare the four routing contexts.
    The left chart reports the shares of traceroutes associated at zero, one,
    or two AS hops from the origin. The right chart reports the shares using
    origin ASes, non-origin ASes, AS links, and ordered AS pairs.}
    \label{fig:path-feature-distributions}
\end{figure}

We analyze the path features selected for 1,179,509 traceroutes that we
associate with a unique origin AS. When multiple feature types tie at
the closest origin distance and the latest observed hop, we retain them all,
so the percentages by feature type can overlap.

\textbf{Features near the origin extend inference beyond direct origin observations.}
As Figure~\ref{fig:path-feature-distributions} (left) shows, we infer the
forwarding direction using features one or two AS hops from the candidate
origin for 60.3\% of identified traceroutes. Notably, we do not directly
observe the candidate origin AS in 99.9\% of these traceroutes. Using path
features near the origin allows us to infer their forwarding directions
while limiting extrapolation to at most two AS hops.

\textbf{Binary path features are particularly useful for distinguishing
forwarding directions when alternatives exist.}
As Figure~\ref{fig:path-feature-distributions} (right) shows, non-origin AS
nodes are selected for 79.3\% of identified traceroutes without alternatives,
compared with 30.1--52.7\% when alternatives exist. One likely reason is that
paths toward invalid and alternative origins can share transit ASes.
Observing a shared AS alone cannot distinguish the two directions, but a
particular AS link or ordered AS pair may appear only on paths toward one
of the origins. In contexts with alternatives, AS links and ordered AS pairs
are selected for 41.7--51.2\% and 73.7--85.5\% of identified traceroutes,
respectively.

\subsection{Recovering Directions from Incomplete Traceroutes}
\label{sec:direction-recovery}

To evaluate the accuracy and recovery capability of our method on incomplete
traceroutes, we construct a ground-truth dataset from well-observed traceroutes.
We select June 30 traceroutes that received responses at more than 80\% of the
hop positions before the last responsive hop and for which all responding IP
addresses outside the target prefix can be mapped to ASes. To label a traceroute,
we require at least one observed candidate origin. Each such origin must
originate routes for only one forwarding direction and be absent from BGP paths
for the other direction. Finally, we obtain 82,084 traceroutes with ground-truth
labels for their observed forwarding directions.

We mask all responses from candidate origin ASes to test whether our method can
infer forwarding directions from the remaining hops. To simulate packet loss
during measurement, we randomly drop the remaining hop responses at loss
probabilities ranging from 0\% to 80\%, with 20 trials per setting. For each routing
context, we examine whether the observed forwarding directions are fully
recovered, partially recovered, or missed, and whether any inferred directions
are absent from the ground-truth labels.

\begin{figure}[!t]
  \centering
  \includegraphics[width=\columnwidth]{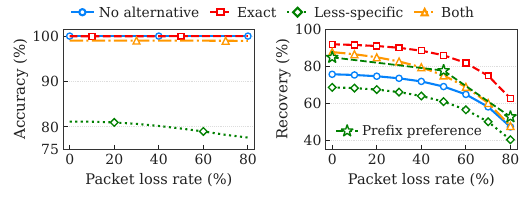}
  \caption{Accuracy (left) and complete recovery (right) under response
    loss after masking origin responses. Stars show results with prefix preference.}
  \Description{Two plots side by side compare accuracy on the left
    and complete recovery on the right across four routing contexts.
    Recovery decreases as packet loss increases, while accuracy among
    inferred results remains comparatively stable.
    Three green stars connected by a dashed line show improved recovery with prefix preference at zero,
    fifty, and eighty percent packet loss.}
  \label{fig:direction-recovery}
\end{figure}

\textbf{Our method accurately infers forwarding directions even when origin
ASes are unobserved.}
After masking origin responses, we find that the inferred directions exactly
match the ground-truth labels for over 99\% of traceroutes with a determined
direction in the no-alternative, exact-prefix, and both-alternatives contexts.
In the less-specific context, the inferred directions still exactly match the
labels for 81.12\% of these traceroutes.
As packet loss increases, accuracy remains around 99\% or higher in these
three contexts. In the less-specific context, it decreases from 81.12\% to
77.60\% at 80\% packet loss.

We find that discrepancies after origin masking stem from two causes.
First, some traceroutes retain discriminative features for a direction absent
from the ground-truth label, leading our method to report that direction. However, the original
traceroute measurements captured responses from only one candidate origin,
so the ground-truth label covers just one direction. This discrepancy therefore reflects
a limitation of ground-truth construction rather than an inference error.
Second, some traceroutes fail to recover a direction in their ground-truth labels
after masking. Our method correctly identifies that direction in the original traceroutes. After origin responses
are removed, however, the remaining observations no longer contain qualifying
discriminative features for that direction. Our method therefore
retains the direction whose features remain available.

\textbf{Our method remains resilient to substantial response loss.}
As shown in Figure~\ref{fig:direction-recovery} (right), complete recovery after
origin masking declines gradually at moderate packet loss rates and more
steeply under severe loss. Across routing contexts, we recover all labeled
directions in 60.85--85.94\% of traceroutes at 50\% packet loss and in
40.22--62.63\% even at 80\% loss. Recovery depends on whether the remaining
observations preserve discriminative path features. When packet loss makes
these features unobservable, the surviving hop responses may no longer provide
enough evidence to identify the forwarding direction.

We observe that recovery is consistently lowest in the less-specific context.
To understand this difference, we investigate further and find that some
unrecovered traceroutes retain features shared by the invalid and less-specific
alternative routes after origin masking but have lost the features needed to distinguish their
forwarding directions. We then preferentially attribute these shared features
to the more-specific invalid route while keeping the same distance threshold.
With this attribution, complete recovery without packet loss increases from
68.57\% to 84.81\%. At packet loss rates of 50\% and 80\%, it increases from
60.85\% to 77.64\% and from 40.22\% to 52.62\%, respectively.

\subsection{Resilience to Control-Plane Data Limitations}
\label{sec:control-plane-robustness}

We demonstrate that our forwarding inferences remain robust to incomplete
control-plane visibility and temporal lag.

\begin{figure}[!t]
    \centering
    \includegraphics[width=0.98\linewidth]{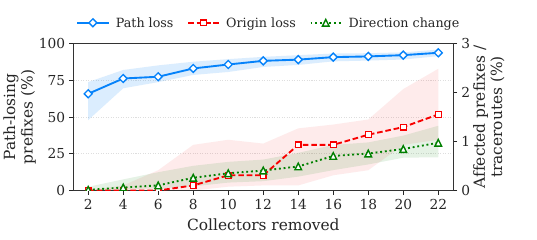}
    \caption{Sensitivity to BGP collector coverage. Lines show medians and
    bands span the 25th to 75th percentiles across trials.}
    \Description{As more BGP collectors are removed, the percentage of
    prefixes losing at least one path rises rapidly, then levels off. The
    percentages of prefixes losing an origin and traceroutes changing
    direction rise more slowly and use a separate vertical scale.}
    \label{fig:collector-sensitivity}
\end{figure}

\textbf{Resilience to limited control-plane visibility.}
Using the June 30 measurement round, we randomly withhold up to 22 of the 53
route collectors, with 100 trials at each tested subset size. For each trial,
we reconstruct routing contexts and path features and compare the inferred
forwarding directions with those obtained using all collectors.
Figure~\ref{fig:collector-sensitivity} shows that as we increase the number of
withheld collectors from 2 to 22, the median percentage of prefixes losing at
least one visible BGP path rises from 66\% to 94\%. Over the same range,
however, the corresponding percentage for loss of at least one visible
candidate origin rises from 0\% to just 1.6\%. Likewise, changes in inferred
forwarding direction remain rare, with the median rising from 0.013\% to
0.975\% of traceroutes and a maximum of 2.36\% across all trials.

\textbf{Resilience to temporal lag.}
To assess sensitivity to con\-trol-plane timeliness, we reconstruct routing
contexts and path features using RouteViews RIB dumps and the corresponding
VRPs from an adjacent time window. This shifts the BGP observation
times by 2--4 hours. The classification changes for only 0.25\% of
reevaluated traceroutes.

\section{Discussion}
\label{sec:discussion}

\textbf{Operational implications.}
An operator's RIB and FIB identify the locally selected route and next hop, but do not reveal
how downstream ASes forward the traffic. As partial ROV deployment persists across the Internet,
our study bridges the gap between local route selection and downstream data-plane forwarding,
uncovering operational blind spots that merit closer scrutiny from network operators.
Operators can collect traceroutes from different PoPs to compare the inferred forwarding
origin with the origin of the locally selected route. Discrepancies can
guide further investigation into downstream route propagation, ROV filtering, and interdomain
routing policies.

\textbf{Measurement limitations and generalizability.}
Our inferences depend on traceroute observations and collector-visible BGP paths.
Section~\ref{sec:association-diagnostics} shows that our method tolerates missing
traceroute responses and limited control-plane visibility. Nevertheless, sufficient
traceroute observations remain necessary to identify discriminative path features.
When these features are no longer observable, the remaining responses provide
insufficient evidence to recover the corresponding forwarding direction. BGP paths
absent from public collectors may also affect which features appear specific to a
candidate origin. IP-to-AS mapping errors and ECMP can alter the observed AS paths
and thereby affect our path-feature matching~\cite{augustin2006paris}.

The path association method does not depend on 6in4 tunnels or IPv6. It can also be applied to
IPv4 and IPv6 traceroutes collected through other distributed measurement infrastructures,
such as RIPE Atlas~\cite{ripeatlas2015}.

\section{Conclusion}

In this paper, we presented a measurement study characterizing data-plane forwarding beyond the
next hop for prefixes with naturally occurring RPKI-invalid IPv6 routes. To overcome incomplete
data-plane paths, we developed a context-aware path association methodology that resolves
forwarding directions using origin-proximal topological features. Our evaluation shows that
forwarding directions remain recoverable even without origin responses and under substantial
response loss, while the inferred results remain stable under limited control-plane visibility
and temporal offsets.

Our measurements reveal that although non-invalid alternatives dominate forwarding when
available, invalid-origin forwarding remains pervasive over time across measured prefixes and
ASes. Longitudinal remeasurements after routes become valid further reveal that many prematurely
terminating traceroutes result from route unavailability. Furthermore, our case analyses show
how ROV enforcement, peering interconnections, and export policies shape the propagation and
forwarding impact of routes left invalid by operational inconsistencies between ROA
authorizations and BGP announcements. Filtering can sharply limit invalid-route propagation,
yet collateral damage can persist under partial ROV deployment.

\bibliographystyle{ACM-Reference-Format}
\bibliography{refs}

\appendix
\section{Ethics}
\label{sec:ethics}

We follow established principles for ethical Internet measurement to minimize the impact on
network operations and third-party infrastructure~\cite{menlo}. We send only ICMPv6 probes
for tunnel endpoint identification and traceroute. All probes carry our measurement server's IPv6
address as the source, so responses return directly to infrastructure under our control. We cap the
aggregate sending rate at 5 MB/s, limit traceroutes to 30 hops, and sample at most three tunnel
endpoints per network. We also distribute probes across endpoints and targets to avoid
concentrating measurement traffic.

\section{RPKI-Invalid Prefix Attribution}
\label{app:attribution}

We adapt the attribution method in~\cite{li2026demystifying} to identify
operational explanations for the RPKI-invalid IPv6 prefix--origin pairs in our
measurements.

\textbf{Datasets.} For the June 30 measurement round, we use BGP routes
from RouteViews~\cite{routeviews} and contemporaneous VRPs from
Routinator~\cite{routinator}. We identify AS organizational affiliations
using CAIDA's AS-to-Organization Mapping Dataset~\cite{caida_as_org} and
IIL-AS2Org~\cite{chen2023siblings,chen2026as2org}, and provider--customer
relationships using CAIDA's AS Relationships Dataset~\cite{caida_as_rel}.
To identify candidate opaque-transit providers and direct lessors, we
analyze RPKI-valid IPv6 routes in RouteViews RIB dumps from November 18,
2025 to June 26, 2026. We use a minimum of $k=3$ distinct potential customer
ASes, balancing coverage and selectivity.

\textbf{Supplementary evidence and forwarding comparison.}
We use public IRR \texttt{route6} objects~\cite{rfc4012} to check for address
delegation and consult PeeringDB~\cite{peeringdb} and CAIDA
ASRank~\cite{caida_asrank} for evidence of AS business roles.

For the comparison in Section~\ref{sec:widespread-invalid-forwarding}, we
include all 956 prefixes measured from at least 20 ASes each. For each prefix,
we calculate the percentage of measured ASes exhibiting invalid-origin
forwarding. Table~\ref{tab:attribution-forwarding} breaks down invalid-origin
forwarding by attribution status, separately for prefixes with and without
alternative routes.

\begin{table}[!hb]
  \centering
  \caption{Invalid-origin forwarding by attribution status.
  Affected AS percentages are averaged across prefixes.}
  \label{tab:attribution-forwarding}
  \footnotesize
  \begin{tabular}{@{}lcccc@{}}
    \toprule
      & \multicolumn{2}{c}{\textbf{Without alternatives}}
      & \multicolumn{2}{c@{}}{\textbf{With alternatives}} \\
    \cmidrule(lr){2-3}\cmidrule(l){4-5}
    \textbf{Attribution}
      & \textbf{Prefixes} & \begin{tabular}[c]{@{}c@{}}\textbf{Affected}\\\textbf{ASes (\%)}\end{tabular}
      & \textbf{Prefixes} & \begin{tabular}[c]{@{}c@{}}\textbf{Affected}\\\textbf{ASes (\%)}\end{tabular} \\
    \midrule
    Full    & 358 & 30.6 & 322 & 4.5 \\
    Partial &   9 &  8.6 &   0 & --- \\
    None    & 159 &  9.7 & 108 & 4.7 \\
    \bottomrule
  \end{tabular}
\end{table}

\section{Temporal Evolution of Competing Exact-Prefix Routes}
\label{app:exact-prefix-evolution}

We examine in more detail how route propagation and forwarding changed for
the prefix \texttt{2001:\allowbreak43f8:1400::/48} discussed in
Section~\ref{sec:widespread-invalid-forwarding}.
We compared RouteViews RIB
dumps from 22:00 UTC on June 16 and 00:00 UTC on June 17. The valid route
to AS328650 via AS5511 contracted from 223 observed AS paths to two, while
the competing invalid route to AS8346 expanded from 26 paths to 36. Among
the ASes that disappeared from valid paths, 11 appeared on newly observed
invalid paths. Other valid routes observed through AS5511 or AS8346 remained
stable.

To trace this change, we inspected BGP updates and identified explicit path
replacements at two of these ASes. At 23:09:12 and 23:09:15 UTC on June 16,
AS24940 and AS32590 replaced their valid paths via AS5511 with invalid
paths via AS6762, within seconds of the first observed withdrawal at
23:09:11 UTC.

Our next measurement captured the corresponding shift in forwarding. From
June 16 to June 17, the percentage of measured ASes forwarding toward the
valid origin fell from 32.1\% to 2.6\%, while the percentage forwarding
toward the invalid origin rose from 39.3\% to 59.5\%. By June 18, none
forwarded toward the valid origin, whereas 61.3\% forwarded toward AS8346.

\section{Forwarding along Invalid and Alternative Routes}
\label{app:multiple-forwarding}

In Section~\ref{sec:origin-direction}, we show that some traceroutes exhibit
forwarding along both invalid and alternative routes. Here we analyze the 46,508
traceroutes with multiple associations across 149 prefixes to determine
which route's features appear first.

\textbf{Identifying forwarding transitions.}
We retain all route-specific matches within two AS hops of a candidate
origin along a BGP path. To determine their order in a traceroute, we use
the position of each matched AS node and both endpoints of each matched
AS link or ordered AS pair. We identify a forwarding transition from
alternative to invalid routes when an alternative-specific feature ends
before any invalid-specific feature begins and the last invalid-specific
feature ends later than the last alternative-specific feature. We reverse
the rule for transitions from invalid to alternative routes. The remaining
traceroutes are left unclassified.

As Table~\ref{tab:multiple-feature-order} shows, transitions predominantly
proceed from alternative to invalid routes with less-specific alternatives,
whereas the reverse predominates with exact-prefix alternatives.

\suppressfloats[t]
\begin{table}[ht]
  \centering
  \footnotesize
  \caption{Order of route-specific evidence in the June 30 measurement round.
  Percentages use all traceroutes matching both invalid and alternative routes
  in each context.}
  \label{tab:multiple-feature-order}
  \begin{tabular}{@{}lcc@{}}
    \toprule
    \textbf{Alternatives} & \textbf{Alternative first} & \textbf{Invalid first} \\
    \midrule
    Less-specific & 78.7\% & 1.2\% \\
    Exact-prefix & 0.5\% & 72.8\% \\
    Both & 40.8\% & 26.6\% \\
    \bottomrule
  \end{tabular}
\end{table}

With less-specific alternatives, we
identify two forwarding scenarios: downstream diversion, as discussed in
Section~\ref{sec:widespread-invalid-forwarding}, and transit through a covering
origin toward the invalid origin. With exact-prefix alternatives, we observe
traffic traversing the invalid origin before continuing toward the valid
origin, either its direct customer or reachable through intermediate
networks. When both types of alternatives coexist, we commonly observe
transitions in both directions. The cases below illustrate how these forwarding
transitions arise across distinct routing contexts.

\subsection{Less-Specific Alternatives}
\label{app:multiple-less-specific}

We illustrate two cases in which traffic switches to a more-specific invalid route.

\begin{figure}[!t]
  \centering
  \includegraphics[width=0.90\linewidth,clip]{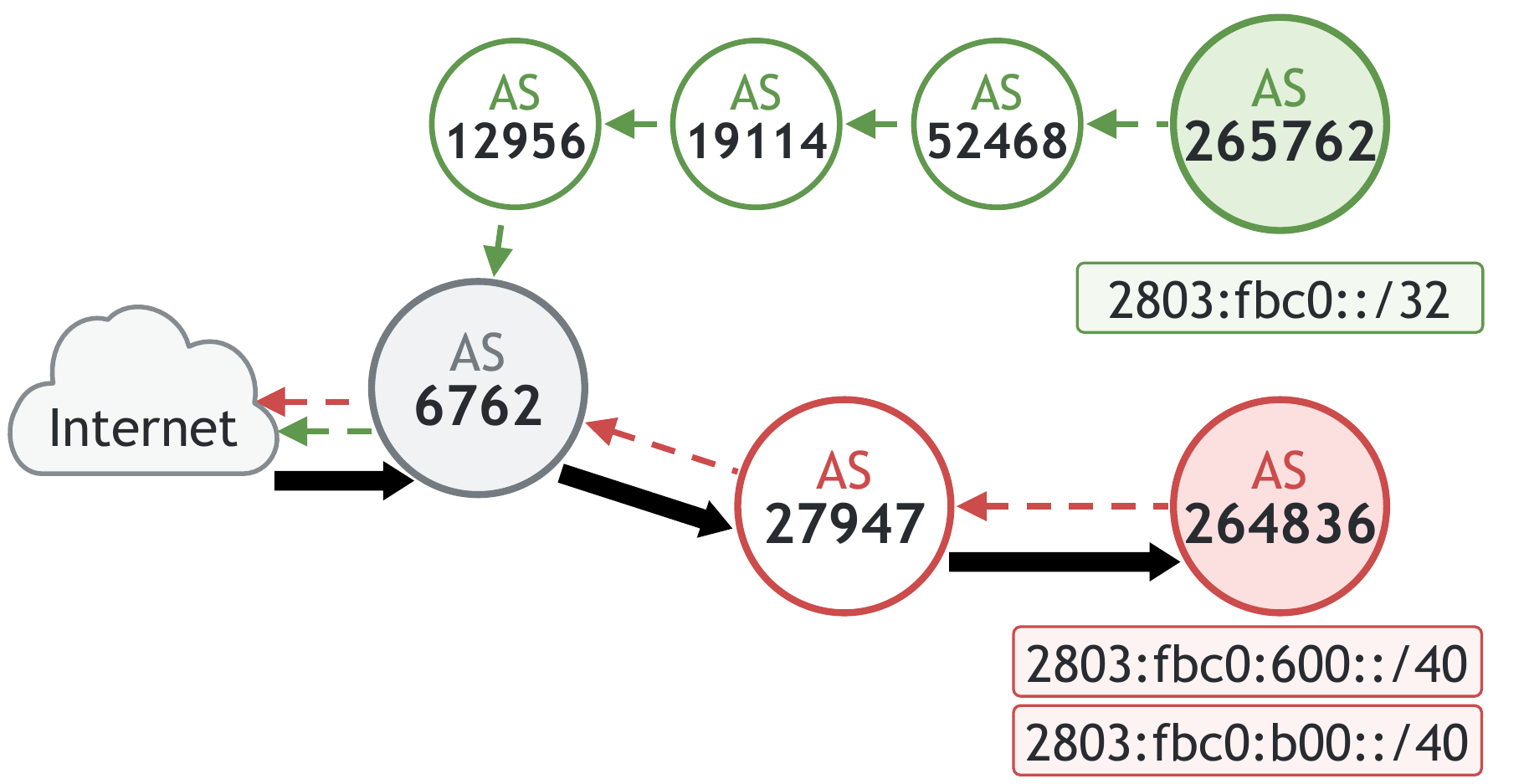}
  \caption{AS6762 diverts traffic onto two invalid /40 routes covered by
  \texttt{2803:fbc0::/32}.}
  \Description{A single-column network diagram shows the Internet and AS6762
  on the left. Green dashed arrows show propagation of the valid covering
  route from AS265762 through AS52468, AS19114, and AS12956. Red dashed arrows
  show propagation of the invalid /40s from AS264836 through AS27947 and
  AS6762. Black solid arrows show traffic forwarded along the invalid route.}
  \label{fig:less-specific-diversion-fbc0}
\end{figure}

\textbf{AS6762 redirects traffic onto more-specific invalid routes.}
For the three prefixes discussed in Section~\ref{sec:widespread-invalid-forwarding},
we examine traceroutes that traverse AS6762 and switch from the covering
route to the invalid route. In 86.8\% of these traceroutes, traffic follows
the covering route into AS6762 and leaves for AS27947 along the more-specific
invalid route. Figure~\ref{fig:less-specific-diversion-fbc0} illustrates this
transition for the two invalid /40s covered by \texttt{2803:fbc0::/32}.

\textbf{Akamai forwards traffic onward to BSNL.}
Akamai (AS20940), a global CDN operator, announces the valid covering
aggregate \texttt{2600:\allowbreak1400::/24}, whereas Bharat Sanchar Nigam
Limited (BSNL, AS9829), an Indian broadband provider, originates the
RPKI-invalid sub-prefix \texttt{2600:\allowbreak1417:\allowbreak62::/48}.
We find that 2,418 of the 4,722 traceroutes to the /48 (51.2\%) pass through
Akamai before reaching BSNL. For example, one traceroute follows the AS path
AS3356--AS20940--AS6453--AS9498--AS9829.

From BSNL's annual report~\cite{bsnl_annual_report_2017_18} and IRR
records~\cite{akamai_bsnl_arin_route6}, we identify an existing CDN caching and
peering relationship between Akamai and BSNL, as well as another /48 within
Akamai's address block registered with BSNL as its origin. Based on these
findings, we suspect that the measured /48 may also have been delegated to BSNL
for CDN deployment. In this scenario, Akamai advertises the
covering aggregate to the Internet, while BSNL's more-specific /48 announcement
provides a route toward the delegated addresses, allowing Akamai to forward
traffic for the delegated addresses to BSNL. This would explain why traffic
reaching Akamai continues toward BSNL, despite BSNL lacking ROA authorization
for the measured /48.

\subsection{Exact-Prefix Alternatives}
\label{app:multiple-exact-prefix}

With exact-prefix alternatives, 72.8\% of traceroutes that show
forwarding along both routes follow the invalid route first.

\textbf{GTT forwards traffic for customer prefixes toward the valid origin.}
For four /48s, including \texttt{2a0f:\allowbreak7d00:\allowbreak32ba::/48},
GTT (AS3257) originates an RPKI-invalid route, while AS200546 originates a
valid route for the same prefix. We observe the \texttt{3257:4000} community
on GTT's announcements for all four prefixes~\cite{ripe_ris}, indicating
that GTT advertises them as customer routes~\cite{gtt_bgp_communities}.
Across the four prefixes, 71.8--79.0\% of traceroutes that show forwarding
along both routes first reach GTT and are then forwarded toward AS200546.
For instance, a traceroute to \texttt{2a0f:\allowbreak7d00:\allowbreak32ba::/48}
ends with the observed AS sequence
\(\mathrm{AS3257}\to\mathrm{AS6939}\to\mathrm{AS60068}\).
The \(\mathrm{AS6939}\to\mathrm{AS60068}\) segment matches the valid BGP path
\(\mathrm{AS6939}\to\mathrm{AS60068}\to\mathrm{AS200546}\), identifying
forwarding toward AS200546.

\textbf{AS3816 forwards traffic directly to its customer.}
AS3816 originates an RPKI-invalid route for
\texttt{2803:\allowbreak1a90:\allowbreak15::/48}, whereas its customer
AS272881 originates a valid route for the same prefix. In 556 of the 1,236
traceroutes to this prefix (45.0\%), traffic passes through AS3816 and is
subsequently forwarded toward AS272881. Notably, 11 of these traceroutes
follow the path \(\mathrm{AS52320}\to\mathrm{AS3816}\to\mathrm{AS272881}\),
with one missing hop between AS52320 and AS3816.
The AS52320--AS3816 link appears only on BGP paths for the invalid route,
indicating that traffic enters AS3816 via that route before being forwarded
to its customer.

\subsection{Both Types of Alternatives}
\label{app:multiple-both-alternatives}

When exact-prefix and covering alternatives coexist, 40.8\% of traceroutes
that show forwarding along an invalid route and an alternative route follow
the alternative route first, while 26.6\% follow the invalid route first.

\textbf{Telconet forwards traffic to its customer over the invalid route.}
For \texttt{2800:\allowbreak2a0:\allowbreak4003::/48}, Telconet (AS27947)
originates valid exact-prefix and covering routes, while its customer
AS264836 originates the RPKI-invalid /48. This prefix accounts for 1,493
of the 1,498 transitions from alternative to invalid routes in this context.
One traceroute follows \(\mathrm{AS1299}\to\mathrm{AS27947}\to\mathrm{AS264836}\).
The incoming link \(\mathrm{AS1299}\to\mathrm{AS27947}\) is specific to the
covering routes, identifying forwarding along a valid covering route into
Telconet.

\textbf{AS18678 forwards traffic to another InterNexa network.}
For \texttt{2800:\allowbreak650:\allowbreak113::/48}, AS18678 (InterNexa
Colombia) originates the RPKI-invalid /48, while AS28032 (InterNexa Peru)
originates a valid route for the same prefix and a covering /32. Of the
669 traceroutes that show forwarding along both an invalid and an
alternative route, 47 follow the invalid route first. A typical path is
\(\mathrm{AS8966}\to\mathrm{AS18678}\to\mathrm{AS28032}\).
The incoming link AS8966--AS18678 is specific to the invalid route,
identifying forwarding along that route into AS18678. The outgoing link
AS18678--AS28032 appears on BGP paths for both valid alternatives.

\section{Forwarding toward Regional Origins}
\label{app:regional-origin-forwarding}

Three regional networks operated by InterNexa originate the same 10
prefixes. The announcements from AS18678 (InterNexa Colombia) and AS262589
(InterNexa Brazil) are RPKI-invalid, while those from AS28032 (InterNexa
Peru) are valid under the corresponding ROAs.

AS18678 is a candidate origin for all 10 prefixes and also appears on BGP
paths toward the other two origins. Observing AS18678 alone therefore
leaves the forwarding direction ambiguous. Our context-aware path
association method distinguishes these directions using the networks and
links through which traffic enters InterNexa.

\textbf{AS18678 (InterNexa Colombia).} We identify forwarding toward
AS18678 only for \texttt{2800:\allowbreak650:\allowbreak140::/48}.
All 103 associated traceroutes enter AS18678 through AS52468 (UFINET),
an incoming link specific to BGP paths for AS18678's announcement.
For the other 9 prefixes, we do not observe its announcements propagating
through other ASes.

\textbf{AS262589 (InterNexa Brazil).} We identify 160 traceroutes toward
AS262589 across 8 prefixes. In 127 of these traceroutes, we observe a direct
link into AS18678 from either AS8966 (Etisalat) in the UAE or AS9498 (Bharti
Airtel) in India. For the corresponding prefixes, these links appear only on
BGP paths toward AS262589.

\textbf{AS28032 (InterNexa Peru).} We identify forwarding toward this valid
origin for all 10 prefixes. The associated traceroutes primarily traverse
global backbones, including AS6939 (Hurricane Electric) and Tier-1 ASes
such as AS174 (Cogent), AS6762 (Sparkle), AS3356 (Level 3), and AS6453
(Tata Communications).

Forwarding toward AS28032 varies with the propagation scope of valid
alternatives. The valid exact-prefix routes for all 10 prefixes have limited
propagation: we observe them on only one AS path, comprising six ASes.
However, for 7 prefixes, AS28032 also announces a valid covering route for
\texttt{2800:\allowbreak650::/32}, whose observed BGP paths span nearly 250 ASes.
This broader propagation coincides with substantially more forwarding toward
AS28032, with 82.5--90.0\% of traceroutes directed toward this origin.

In contrast, for the remaining 3 prefixes, the valid
\texttt{2803:\allowbreak25e0::/32} route, which exactly matches one prefix and
covers the other two /36s, is also confined to this single AS path. With no
broadly propagated alternative, forwarding toward AS28032 is rare
(0.3--2.5\% of traceroutes), while 88.7--93.2\% remain unresolved. Among these
unresolved traces, 25.3\% terminate with an ICMPv6
\emph{No route to destination} response.

\end{document}